\documentstyle{laa}  

\begin{document}

 \thesaurus{
	03(03.13.6;	
	11.04.1)	
	}

 \title{About the Malmquist bias in the determination of $H_{0}$\\
and of distances of galaxies}

 \author{
	R.~Triay \inst{1}\thanks{Universit\'e de Provence and the
European Cosmological Network}
	 \and M.~Lachi\`eze-Rey \inst{2}\thanks{the European
Cosmological Network}
	 \and S.~Rauzy \inst{3}\thanks{on leave from CPT as Boursier
MRT at Univ. Aix-Marseille II, and
SAP-CEN Saclay}}

\offprints{R.~Triay}

 \institute{
	Centre de Physique Th\'eorique - C.N.R.S., Luminy Case 907, F-13288
Marseille
 Cedex 9, France
 \and
	Service d'Astrophysique, C.E. Saclay, F-91191 Gif sur Yvette Cedex, France
 \and
	Dept. of Physics, Queen's University at Kingston, Canada
	 }

 \date{Received 13 July 1993 / Accepted }

 \maketitle

 \begin{abstract}
We provide the mathematical framework which elucidates the way of
using a Tully-Fisher (TF) like
relation in the determination of the Hubble constant $H_0$, as
well as for distances of
galaxies. The first step toward the comprehension of this problem
is to define a statistical
model which accounts for the (linear) correlation between the
absolute magnitude $M$ and the
line width distance estimator $p$ of galaxies, as it is observed.
Herein, we assume that $M=a.p+b-\zeta$, where $\zeta$ is a random
variable of zero mean describing an intrinsic
scatter, regardless of measurement errors. The second step is to understand
that the
calibration of this law is not unique, since it depends on the statistical
model used for
describing the distribution of variables (involved in the calculations). With
this in mind, the
methods related to the so-called Direct and Inverse TF Relations (herein DTF
and ITF) are
interpreted as maximum likelihood statistics. We show that, as long as the same
model is used
for the calibration of the TF relation and for the determination of $H_0$, we
obtain a coherent
Hubble's constant. In other words, the $H_0$ estimates are not model dependent,
while the TF
relation coefficients are. The choice of the model is motivated by reasons of
robustness of
statistics, it depends on selection effects in observation which are present in
the sample. For
example, if $p$-selection effects are absent then it is more convenient to use
a (newly
defined) robust statistic, herein denoted by ITF$^\star$. This statistic does
not require
hypotheses on the luminosity distribution function and on the spatial
distribution of sources,
and it is still valid when the sample is not complete. Similarly, the general
above results
apply also to distance estimates of galaxies. The difference on the distance
estimates when
using either the ITF or the DTF model is only due to random fluctuations. It is
interesting to point
out that the DTF estimate does not depend on the luminosity distribution of
sources. Both
statistics show a correction for a bias, inadequately believed to be of
Malmquist type. The
repercussion of measurement errors, and additional selection effects are also
analyzed.

\keywords{galaxies : distances and redshifts -- methods : statistical}

\end{abstract}

\section{Introduction}\label{Introduction}
Herein, we regard the Tully-Fisher (TF) relation for spiral galaxies in
optical, Tully \& Fisher
(\cite{TulFis77}), and the Faber-Jackson relation for E galaxies, Faber \&
Jackson (\cite{FabJac76}),
as a single law providing us with an estimate of the absolute magnitude $M = ap
+ b$, where $p$ is
called {\it line width distance estimator}. The determination of the Hubble
constant $H_{0}$, when
a line width distance estimator is involved has long been discussed with
respect to the Malmquist
bias by different authors without reaching yet a general agreement (see e.g.,
Bottinelli et al.
\cite{BotEtal86a},\cite{BotEtal86b},\cite{BotEtal88a},\cite{BotEtal88b};
Giraud \cite{Gir85},\cite{Gir87}; Gouguenheim et al. \cite{GouEtal89};
Jacoby et al. \cite{JacEtal88}; Lynden-Bell et al. \cite{LynEtal88};
Pierce \& Tully \cite{PieTul88}; Tammann \cite{Tam87},\cite{Tam88}; Sandage
\cite{San88a},\cite{San88b}; Teerikorpi
\cite{Tee75},\cite{Tee82},\cite{Tee84},\cite{Tee87},\cite{Tee90}; Tully
\cite{Tul88}). The aim of this work is to enlighten on this problem
through a theoretical point of view, and to provide us with rigorous
formulas for ongoing applications. In this first paper, we seek a
mathematical framework which gives fair rudiments for discussing on the
use of the {\it Direct} Tully-Fisher ({\rm DTF}) relation and the {\it
Inverse} Tully-Fisher ({\rm ITF}) relation, which both interpret as a
choice of a fitting technique.

According to Bigot \& Triay (\cite{BigTri90b}), one must keep in mind that a
technique of fitting is
intimately related to a statistical model. Namely, the related statistics (or
estimates) are
warranted as long as the values distribution of variables involved in the
calculation is correctly
described by such a model. Hence, we understand that a statistical model is as
a matter of fact
required for arguing on the use of either the {\rm DTF} relation or the {\rm
ITF} relation. However,
it must be noted that if a statistical model is available then nothing prevents
us to use solely the
{\it maximum likelihood} ({\sc ml}) technique. Such an approach has the
advantage of providing us
unambiguously with a unique fitting technique. Moreover, the related statistics
give unbiased
estimates of model parameters, as long as the statistical model takes into
account the selection
effects.

Therefore, according to above precepts, in Section\,\ref{Model}, we define the
statistical model
which describes the distribution of variables involved in the determination of
$H_{0}$. In
Section\,\ref{fitting}, we derive the statistics used for the calibration of
the $M$--$p$ relation
and the estimation of $H_{0}$. The influence of selection effects on these
estimates is also
analyzed. In Section\,\ref{Errors}, we investigate the repercussion of
measurement errors.
Section\,\,\ref{DistancesOfGalaxies} enlightens on the definition of a reliable
distance estimate of
galaxies. It is strongly recommended to read the notations and useful formulas
given in
Appendix\,\ref{Notations}, these features are addressed throughout the text by
means of symbol ``{\it
Def.}''.

\section{The basic model}\label{Model}
Herein, we specify the probability density ({\it pd}) describing the
distribution of variables
involved in the calculation. These variables are related to intrinsic
quantities of sources
(galaxies), which are~:
\begin{itemize}
\item the absolute magnitude $M$,
\item the line width distance estimator $p$,
\item the distance $r$ from the observer,
\item and the radial velocity $v$ (corrected for solar motion).
\end{itemize}
For reasons of simplicity in calculations, instead of using the variables $r$
and $v$, it is
more convenient to use the distance modulus
 \begin{equation}\label{ram}
\mu = 25 + 5\log r,
 \end{equation}
where $r$ is given in Mpc, and a similarly defined quantity
 \begin{equation}\label{eta}
\eta = 25 + 5\log v.
 \end{equation}

If the peculiar velocities of sources are neglected then the Hubble law reads
 \begin{equation}\label{HubbleL}
	\eta = \mu + {\cal H},
 \end{equation}
where ${\cal H} = 5\log H_0$. This equation shows that the variables $\mu$ and
$\eta$ define the same
quantity\footnote{This rough description suffices for the present scope,
although this model is
refined in Triay et\,al.\,(\cite{TriEtal93c}).}. If no evolutionary effect of
sources is present then
the distribution of intrinsic quantities $M$ and $p$ are independent on $\mu$
(since a distance
corresponds to a time-shift). Thus, regardless of capacities in observation,
the theoretical {\it pd}
describing the distribution of above variables can be written as follows
 \begin{equation}\label{BasicTh}
	dP_{\rm th} = F(M,p)dMdp~\kappa(\mu)d\mu,
 \end{equation}
see ({\it Def.}1), where $\kappa(\mu)$ accounts for the distribution of
galaxies
in space and $F(M,p)$ for the $M$--$p$ distribution, i.e., the TF diagram. The
projection of the TF\,{\it pd} onto the $M$-axis, resp., the $p$-axis, provides
us with the {\it luminosity distribution function},
 \begin{equation}\label{LuminosityFunction}
	f_{M}(M;M_{0},\sigma_{M}) = \int F(M,p)dp,
 \end{equation}
resp., a {\it pd} function ({\it pdf}) describing the
distribution of $p$'s values,
 \begin{equation}\label{pFunction}
	f_{p}(p;p_{0},\sigma_{p}) = \int F(M,p)dM,
 \end{equation}
see ({\it Def.}1).

It is obvious that this statistical model, as defined by Eq.\,(\ref{BasicTh}),
must be improved for
taking into account observational and selection effects (e.g., the sampling
rules). The difficulty in
detecting faint galaxies involves necessarily the apparent magnitude
 \begin{equation}\label{ApparentMagnitude}
	m = M + \mu.
\end{equation}
If the selection effects depend solely on the apparent magnitude, then the data
distribution is
defined by the following {\it pd}
 \begin{equation}\label{BasicObs}
	dP_{\rm obs} = \frac{\phi_{m}(m)}{P_{\rm th}(\phi_{m})}~dP_{\rm th},
 \end{equation}
where $\phi_{m}(m)$ is called {\it selection function}\footnote{It is a
positive function which works
as a filter response ($0\leq\phi_{\rm m}(m)\leq1$) with respect to the apparent
magnitude.}, and
$P_{\rm th}(\phi_{m})$ is a normalization factor, see ({\it Def.}3).

\subsection{Working hypotheses} \label{WorkingHypotheses}
In order to achieve the statistical model, as defined by Eq.\,(\ref{BasicObs}),
we must
specify the functions $\phi_{m}(m)$, $\kappa(\mu)$ and $F(M,p)$. It turns out
that some
results can be obtained though without a full description of the model, which
provides
an interesting feature for related statistics~:~({\it robustness}).

Since the present scope of our analysis is to enlighten on the problem of
biases, we limit on
formulating simple hypotheses but sufficiently complete. This has the advantage
of avoiding cumbersome
calculations but providing us with fair rudiments common to real situations. A
more realistic
model is given in Triay et\,al.\,(\cite{TriEtal93a}). In the following, for
clearness in
understanding, we apply ourself to specify always the working hypotheses used
for each
step. In general, the standard working hypotheses assume~:
\begin{itemize}
\item[$\bullet h_{1})$]
{\it A magnitude limited complete sample}. This property means that the
selection effect
limits to a cutoff at a given limiting magnitude, herein designated by $m_{\rm
lim}$. Thus, the
selection function reads
 \begin{equation}\label{MagnitudeLimited}
 \phi_{m}(m) = \theta(m_{\rm lim}-m),
 \end{equation}
where $\theta$ is the Heaveside distribution function.
\end{itemize}
It is obvious that
 \begin{equation}\label{LimitingMagnitudePractice}
 m_{\rm lim}\geq\max_{k=1,N} \left\{m_{k}\right\},
\end{equation}
and (in practice) a possible choice is to assume the equality.
\begin{itemize}
\item[$\bullet h_{2})$] {\it A uniform spatial distribution of sources}. The
{\it pdf} of the
distance modulus $\mu$, see Eq.\,(\ref{ram}), related to a uniform distribution
in an Euclidian
space is given by
 \begin{equation}\label{uniform}
 \kappa(\mu) \propto \exp(\beta \mu), \quad {\rm where} \quad
 \beta=\frac{3\ln10}{5}.
 \end{equation}
\end{itemize}
Let us mention that, while we focus on a uniform spatial distribution of
sources, the following
calculations and results are still valid for power law distribution
$\beta\neq\frac{3\ln10}{5}$. In
order to specify the TF\,{\it pdf} $F(M,p)$, we must describe accurately the
TF\,diagram, i.e., the
relation $p \mapsto M = \widetilde{M}(p)$. The observations show that the data
are distributed about
a straight line (which visualizes the TF\,relation), that we denote
$(\Delta_{\rm TF})$. Thus this
line is defined by equation
 \begin{equation}\label{Rough}
\widetilde{M}(p) = a.p + b.
 \end{equation}
In addition to scatter from measurement errors, it is sensible to assume that
an {\it
intrinsic} dispersion is also present\footnote{The practical argument in favour
of this
approach is that presently the TF diagram is still continuously improved by
shrewd
corrections (as the galaxy orientation with respect to the line of sight, the
linear
dimension of the galaxy region from which the line width is measured,
etc\ldots), and
thus that the $M$\,estimate given in Eq.\,(\ref{Rough}) is approximate.}. From
a theoretical
point of view, this might be interpreted as either a lack of exact definition
of the
variable $p$ which accounts for a linear relation, and/or the physics of
galaxies
requires actually additional variables for providing the absolute magnitude
(i.e., the
randomness interprets as the effect of these missing variables). Hence, we
easily
understand that a specific approach for defining $F(M,p)$ would require to
presume a
priori the related (unique) physical process, which is unfortunately not yet
known\footnote{One guesses that this lack of information is mainly responsible
for the
present debate on the choice between the DTF and ITF relations, an issue that
is
clarified below.}.

It must be noted that the goal is not to fit the data to such a model but to
derive the
Hubble Constant $H_{0}$. Hence, in order to perform the {\sc ml} technique, we
ask whether we may
substitute $F(M,p)$ by a suitable function which imitates it in reproducing the
$M$-$p$
correlation. The next section shows that the usual approaches, which consists
of using the {\rm
DTF} and the {\rm ITF} relations, interpret as a matter of fact as a particular
choice of such a
function.

It is advantageous to express a ``linear correlation'' by using a random
variable of
\underline{zero mean}, which accounts for the data dispersion about the
straight line
$(\Delta_{\rm TF})$. In this purpose, it seems natural to use the ``regular''
distance,
which is given by the segment orthogonal to this line. However, with our
theoretical
approach, it is equivalent (and suggested by arguments of simplicity) to use
the following
random variable
 \begin{equation}\label{TFrelation}
	\zeta = \widetilde{M} - M,
 \end{equation}
where $\widetilde{M}$ is given in Eq.\,(\ref{Rough}), which is proportional to
the regular distance (which reads $\zeta.\cos(\arctan a)$). According to
Eq.\,(\ref{Rough},\ref{TFrelation}), obvious calculations show that, regardless
of selection
effects, the standard deviation $\sigma_{\zeta}$ of the $\zeta$-distribution
verifies
 \begin{equation}\label{SigmaZetaMagnitude}
\sigma_{\zeta}^{2}=\left(a\sigma_{p}-\rho_{\rm
 th}(p,M)\sigma_{M}\right)^{2}+\left(1-\rho^{2}_{\rm
th}(p,M)\right)\sigma_{M}^{2},
 \end{equation}
where $\rho_{\rm th}(p,M)$ denotes the theoretical correlation coefficient, see
({\it
Def.}5). It is clear that the existence of an efficient correlation is
expressed by the
following inequality \begin{equation}\label{SigmaZetaInequal} \gamma =
\frac{\sigma_{\zeta}}{\sigma_{M}} \ll 1,
 \end{equation}
or equivalently by $\sigma_{\zeta} \ll
\left|a\right|\sigma_{p}$.
Namely, Eq.\,(\ref{SigmaZetaInequal}) insures that $p$ provides a good estimate
of
the absolute magnitude $M$ from Eq.\,(\ref{Rough}).

In order to proceed with the {\sc ml} technique, we must presume the form of
the $\zeta$-{\it pdf},
hereafter denoted by $g(\zeta;0,\sigma_{\zeta})$, see ({\it Def.}1), which
mimics the data
distribution. In practice, a simple regression analysis should help us to guess
the candidate form to
be used\footnote{It is obvious that it must not be exponential, in order to
ensure an effective
$M$--$p$ correlation.}. Now, a second random variable is required for
specifying entirely the data
distribution, let be $\xi$. For a trusty description of the TF diagram, the
choice of $\xi$ should be
suggested by the appearance of the $M$--$p$ distribution. Thus, in absence of
such information, our
working hypotheses for describing the $M$--$p$ correlation are~:
\begin{itemize} \item[$\bullet h_{3})$] the TF diagram can be mimic
by the {\it pdf}
 \begin{equation}\label{TFClass}
f_{\xi}(\xi)d\xi~ g(\zeta;0,\sigma_{\zeta})d\zeta \approx F(M,p)dMdp,
 \end{equation}
with a Gaussian $\zeta$-distribution
 \begin{equation}\label{ZetaGaussian}
g(\zeta;0,\sigma_{\zeta})=g_{\rm G}(\zeta;0,\sigma_{\zeta}),
 \end{equation}
see ({\it Def.}1.a).
 \end{itemize}

The model defined by the {\it pdf} given by Eq.\,(\ref{TFClass}) is general
enough to
interpret the ITF and DTF methods used in the literature. Namely, the {\rm ITF}
relation
$p=a_{\cal I}M+b_{\cal I}$, and the {\rm DTF} relation $M=a_{\cal D}p+b_{\cal
D}$, correspond to
the following identifications~: \begin{equation}\label{Xi}
\xi= \left\{ \begin{tabular}{ll}
$M$ & in the ITF model\\
$p$ & in the DTF model.
\end{tabular} \right.
\end{equation}
Let us mention that such a definition has the advantage of avoiding a confusion
which is
inherent to usual approaches, e.g., see Teerikorpi\,(\cite{Tee90}). Indeed,
Eq.\,(\ref{TFClass})
tells us that $\xi$ and $\zeta$ are uncorrelated outcomes, thus the use of
conditional
probabilities allows us to estimate a value of $p$ from a value of $M=\xi$, in
the case
of ITF model, and we have the reverse situation in the case of DTF model, see
Eq.\,(\ref{Rough},\ref{TFrelation}). On the other hand, as long as the related
random
processes are not specified, the usual definitions would wrongly suggest that
$a_{\cal
I}=1/a_{\cal D}$ and $b_{\cal I}=-b_{\cal D}/a_{\cal D}$.

\section{The technique of fitting}\label{fitting}
In the following, we proceed as follows~: for each model, as defined by
Eq.\,(\ref{TFClass}-\ref{Xi}), we investigate the {\it calibration} of the
TF~relation ({\it
Step}\,1), see Eq.\,(\ref{Rough}), and the {\it determination of $H_{0}$} ({\it
Step}\,2).

In order to establish the likelihood function, the {\it pdf} is written in
terms of
observables, see ({\it Def.}4). The definition of these variables depends on
the step of the
analysis~: \begin{itemize}
\item {\it Step}\,1\,) For the calibration of the TF relation, the data sample
corresponds to
the following observables
 \begin{equation}\label{DataCalibration}
	\left\{ \mu_k, p_k, M_k = m_k - \mu_k \right\}_{k=1,N_{1}},
\end{equation}
see Eq.\,(\ref{ApparentMagnitude}).
\item {\it Step}\,2\,) For the determination of the Hubble constant, it is more
convenient to use the following observables
 \begin{eqnarray}
	\label{variableX} x &=& m - \eta ,\\
	\label{variableY} y &=& a.p + b + \eta - m ,
 \end{eqnarray}
see Eq.\,(\ref{HubbleL},\ref{ApparentMagnitude}), and the data sample
corresponds to
\begin{equation}\label{DataDetermination}
\left\{\eta_k, x_k, y_k \right\}_{k=1,N_{2}}.
\end{equation}
\end{itemize}
We use the following notations~: $N_{1}$, for the sample size, $\langle
\rangle_{1}$, for
the sample average, ${\rm Cov}_{1}$, for the sample covariance, etc\ldots, in
{\it Step}\,1,
while $N_{2}$, $\langle \rangle_{2}$, ${\rm Cov}_{2}$,etc\ldots, in {\it
Step}\,2, see ({\it
Def.}5), which helps us to disentangle the statistics involved in each step. We
must be aware
that the random variables $y$ and $\zeta$ (thus $\sigma_{\zeta}$) are model
dependent
(through the estimates of $a$ and $b$), while they are uniquely defined by
Eq.\,(\ref{Rough},\ref{TFrelation},\ref{variableY}).

\subsection{Regardless of the Tully-Fisher relation}\label{NoTFRelation}
For reasons that appear clear in the following, let us derive the {\sc ml}
estimator of
the Hubble constant when the TF-relation is ignored. The related statistical
model is
obtained by integrating the {\it pd} given by Eq.\,(\ref{BasicTh}) over the
variable $p$.
According to Eq.\,(\ref{LuminosityFunction}), we obtain a theoretical {\it pd}
which reads
 \begin{equation}\label{BasicThC}
	\propto f_{M}(M;M_{0},\sigma_{M})dM~\kappa(\mu)d\mu.
 \end{equation}
By writing this {\it pd} in terms of observables $x$ and $\eta$, see
Eq.\,(\ref{eta},\ref{variableX}), we easily understand that, for deriving an
$H_{0}$\,estimator, we must specify the {\it pdf}s $f_{M}(M;M_{0},\sigma_{M})$
and
$\kappa(\mu)$. Let us assume~:
 \begin{itemize}
\item[$\bullet h_{4})$] a Gaussian luminosity distribution function,
 \begin{equation}\label{MGaussian}
 f_{M}(M;M_{0},\sigma_{M})=g_{\rm G}(M;M_{0},\sigma_{M});
 \end{equation}
\end{itemize}
and hypothesis ($h_{2}$), i.e., a uniformly spatial distribution, see
Eq.\,(\ref{uniform}). Hence, the {\it lf} is given by
 \begin{equation}
 \label{LikelyFunctionITFClass}
	{\cal L}(H_{0}) = -\ln P_{\rm th}(\phi_{m}) - \beta {\cal H} -
\frac{1}{N_{2}}\sum_{k=1}^{N_{2}}\frac{(x_{k} - M_{0} + {\cal
H})^2}{2(\sigma_{M})^2 },
 \end{equation}
see ({\it Def.}4.b). Obvious calculations show that the normalization factor
does not depend on $H_{0}$ as long as the selection effects are free of
velocity
criteria. Hence, the {\sc ml} equation provides us with the following statistic

\begin{equation}\label{H0InverseC}
		{\cal H}^{\rm C} = \left(M_{0}-\beta\sigma_{M}^2\right) -
 \langle x \rangle_{2},
\end{equation}
independently on whether the sample is complete up to a limiting magnitude. It
is
important to mention that, since $\partial P_{\rm th}(\phi_{m})/\partial {\cal
H} = 0$,
the term $\beta\sigma_{M}^2$ in Eq.\,(\ref{H0InverseC}) does not interpret as a
Malmquist
bias correction\footnote{While the sample average of absolute magnitudes in a
magnitude
limited sample is indeed biased because of the Malmquist effect, it is given by
$\langle
M \rangle_{2}=M_{0}-\beta\sigma_{M}^2$ under hypotheses
($h_{1}$,$h_{2}$,$h_{4}$).}, see
({\it Def.}6). In the case of $\eta$-selection effects, it is easy to show that
the
estimator (\ref{H0InverseC}) transforms by substituting $\beta$ by $\beta +
\partial \ln
P_{\rm th}(\phi_{m}\phi_{\eta})/\partial{\cal H}$, where $\phi_{\eta}$ is the
selection
function describing the selection effects on velocities. In order to calculate
the
accuracy of Eq.\,(\ref{H0InverseC}), we need to specify the form of the
selection
function. For a magnitude limited complete sample, i.e., ($h_{1}$), see
Eq.\,(\ref{MagnitudeLimited}), the $x$-distribution reads $\propto g_{\rm
G}(x;M_{0}-{\cal H}-\beta(\sigma_{M})^2,\sigma_{M})$, which shows that the
standard
deviation is equal to
\begin{equation}\label{DiscreteAccuracy} \sigma_{{\cal H}^{C}} =
\frac{\sigma_{M}}{\sqrt{N_{2}}}.
\end{equation}
It is obvious that more accuracy is expected when the TF\,relation is used.

\subsection{The inverse Tully-Fisher relation}\label{InverseTF}
Now we take into account the TF\,relation, by using the {\it pdf} defined by
Eq.\,(\ref{TFClass}-\ref{Xi}). The ITF\,model is specified by the choice $\xi =
M$, which
means that the random variables $M$ and $\zeta$ are not
correlated\footnote{Actually,
these variables may not be necessarily independent, e.g., the $\zeta$-{\it pdf}
may be
Gaussian with a $M$-dependent standard deviation,
$g\left(\zeta;0,\sigma_{\zeta}(M)\right)$, while
Eq.(\ref{LuminosityFunction},\ref{InverseTFDensity}) are still fulfilled.}.
Namely, the
data distribution in the $M$--$p$ plane is supposed to be described by the
following {\it
pd}
 \begin{equation}\label{InverseTFDensity}
	f_{M}(M;M_{0},\sigma_{M})dM g_{\rm G}(\zeta;0,\sigma_{\zeta})d\zeta \approx
 F(M,p)dMdp.
 \end{equation}
Let us remind that the precise rule of this {\it pd} is to mimic the data
distribution, without interpreting the physical process involved in the
TF\,diagram. This
explains the respective locations of terms with respect to the equal sign in
Eq.\,(\ref{InverseTFDensity}). The calculations, given in
Appendix\,\ref{CalculationsITF},
provide us with the following results~:

{\it Step}\,1) The {\sc ml} equations yield statistics of $a \approx a^{\rm
ITF}$, $b \approx b^{\rm ITF}$ and $\sigma_{\zeta} \approx
\sigma_{\zeta}^{\rm ITF}$, which are defined as follows
 \begin{eqnarray}
 \label{aITF} a^{\rm ITF}&=&\frac{(\Sigma_{1}(M))^2}{{\rm Cov}_{1}(p,M)},\\
 \label{bITF} b^{\rm ITF}&=&\langle M \rangle_{1} - a^{\rm ITF} \langle p
 \rangle_{1}, \\
 \label{zetaITF} \sigma_{\zeta}^{\rm ITF}&=&\Sigma_{1}(M)
 \sqrt{\frac{1}{\rho_1^2(p,M)}-1 }.
 \end{eqnarray}
It is interesting to note that, regarded as conventional estimators, these
statistics show
no correction for the Malmquist bias. Let us emphasize that they are valid for
any form of
the selection function $\phi_{m}$. We easily understand that such a feature is
of
particular interest because a smooth decreasing function describes the
selection effects
on apparent magnitude more realistically than a sharp cutoff, as it is assumed
by
hypothesis ($h_{1}$). Moreover, because of the same reasons, it turns out that
these
statistics still work whatever the forms of functions $f_{M}$ and $\kappa$,
i.e., for
any type of luminosity and spatial distributions of sources\footnote{Note that
the {\it
pdf} $\kappa$ might simultaneously describe the selection effects on
distance.}. The
(mathematical) reason of such properties is that the normalization factor
$P_{\rm
th}(\phi)$ does not depend on model parameters $a$, $b$ and $\sigma_{\zeta}$,
which is
the case when the selection function reads $\phi=\phi_{m}$ or
$\phi=\phi_{m}~\phi_{\mu}$,
see Appendix\,\ref{CalculationsITF}.

{\it Step}\,2) It turns out that the form of the ITF\,{\it pdf}, as given by
Eq.\,(\ref{InverseTFDensity}) (see also Eq.\,(\ref{BasicDet})), allows us to
derive
straightforwardly a first estimator, which is given by
\begin{equation}\label{H0InverseS}
		{\cal H}^{\rm ITF^\star} = \langle y \rangle_{2}.
 \end{equation}
It provides us with $H_{0}$ within the standard deviation
\begin{equation}\label{ITFSAccuracy}
\sigma_{{\cal H}^{\rm ITF^\star}} = \frac{\sigma_{\zeta}^{\rm
 ITF}}{\sqrt{N_{2}}},
\end{equation}
see Eq.\,(\ref{zetaITF}). Let us emphasize that it is obtained by preserving
the above
advantages, i.e., without assumptions on the completeness of the sample, the
spatial and
luminosity distributions.

On the other hand, the derivation of the {\sc ml} statistics forces us to
specify the
functions $\kappa$ and $f_{M}$. Thus, in addition of hypotheses
($h_{2}$,$h_{3}$), see
Eq.\,(\ref{uniform},\ref{TFClass}), we assume a Gaussian luminosity
distribution function
($h_{4}$), see Eq.\,(\ref{MGaussian}). Hence, we obtain the following
$H_{0}$\,statistic
\begin{equation}\label{H0Inverse}
		{\cal H}^{\rm ITF} =
 \frac{{\cal H}^{\rm ITF^\star}
+ \gamma^{2}{\cal H}^{\rm C}}
 {1+\gamma^{2}},
 \end{equation}
where $\gamma=\gamma^{\rm ITF}$, see
Eq.\,(\ref{SigmaZetaInequal},\ref{H0InverseC},\ref{H0InverseS}). The accuracy
of such an
estimate is calculated by specifying the function $\phi_{m}$. By assuming
($h_{1}$), we
obtain a standard deviation of \begin{equation}\label{ITFAccuracy}
\sigma_{{\cal H}^{\rm
ITF}} = \frac{\sigma_{{\cal H}^{\rm ITF^\star}}}{\sqrt{1+\gamma^{2}}}.
\end{equation}
It is obvious that Eq.\,(\ref{H0Inverse}) can be interpreted as a weighted mean
of
$H_{0}$\,estimators, where the weighting factors correspond to related
accuracies.
Equation\,(\ref{ITFAccuracy}) shows that the ITF estimate is more accurate
than the ITF$^\star$ one. Hence, we understand that the advantage of having
less
constraints on the validity domain of the ITF$^\star$\,estimator (i.e., to have
weak
working hypotheses for defining the ITF$^\star$ model), is to the detriment of
the
accuracy of estimates.

It turns out that both estimators ${\cal H}^{\rm ITF}$ and ${\cal H}^{\rm
ITF^\star}$ can
be used even when the sources are selected upon velocity criteria. The reason
is that
the related selection function $\phi_{m}(x+\eta)\phi_{v}(\eta)$,
does not disturb the independence of $y$ with respect to variables $\eta$
and $x$, see Eq.\,(\ref{BasicDet}). On the other hand, these statistics
become ineffective when selection rules are based on $p$, because $P_{\rm
th}(\phi_{m}~\phi_{p})$ depends as a matter of fact on $a$ and $b$.

It is interesting to note that little algebra allows us to write
Eq.\,(\ref{H0Inverse}) as
\begin{equation}\label{H0InverseRem}
		{\cal H}^{\rm ITF} = \frac{
 \left(\langle y \rangle_{2} - \beta\left(\sigma_{\zeta}^{\rm
ITF}\right)^{2}\right)
+ \gamma^{2}\left(M_{0} -
 \langle x \rangle_{2}\right)}
 {1+\gamma^{2}}.
 \end{equation}
While providing the same quantity as Eq.\,(\ref{H0Inverse}),
Eq.\,(\ref{H0InverseRem}) is
a different weighted mean of two quantities which are not $H_{0}$\,estimators.
The interpretation of this formulae is enlightened in
Section\,\ref{DistancesOfGalaxies}.

\subsection{The (direct) TF relation}\label{DirectTF}
The underlying working hypothesis used in the {\rm DTF} approach is that
the random variables $p$ and $\zeta$ are not correlated. Namely, the data
distribution in the $M$--$p$ plane is supposed to be described by the
following {\it pd}
 \begin{equation}\label{DirectTFDensity}
 	f_{p}(p;p_{0},\sigma_{p})dp ~g_{\rm G}(\zeta;0,\sigma_{\zeta})d\zeta
\approx
 F(M,p)dMdp.
 \end{equation}
It turns out that this approach forces us to presume a priori the form of
functions $\phi_{m}(m)$, $f_{p}(p;p_{0},\sigma_{p})$ and $\kappa(\mu)$. Hence,
we use
($h_{1}$,$h_{2}$,$h_{3}$), see
Eq.\,(\ref{MagnitudeLimited},\ref{uniform},\ref{TFClass}), and
for reason of coherence with $h_{4}$, we assume~:\begin{itemize}
\item[$\bullet h_{4}^\prime)$] a Gaussian $p$-distribution,
 \begin{equation}\label{pdistribution}
 f_{p}(p;p_{0},\sigma_{p})=g_{\rm G}(p;p_{0},\sigma_{p}).
 \end{equation}
\end{itemize}
The calculations, which are given in Appendix\,\ref{CalculationsDTF}, provide
us with the
following results~:

{\it Step}\,1) The likelihood equations yield
statistics of $a \approx a^{\rm DTF}$, $b \approx b^{\rm DTF}$ and
$\sigma_{\zeta} \approx
\sigma_{\zeta}^{\rm DTF}$, which are defined as follows~:
\begin{eqnarray}
\label{aDTF} a^{\rm DTF}&=&\frac{{\rm Cov}_{1}(p,M)}{(\Sigma_{1}(p))^{2}},\\
\label{bDTF} b^{\rm DTF}&=&\langle M \rangle_{1} - a^{\rm DTF} \langle p
 \rangle_{1}+
\beta \left(\sigma_{\zeta}^{\rm DTF}\right)^{2}, \\
\label{zetaDTF} \sigma_{\zeta}^{\rm
DTF}&=&\Sigma_{1}(M) \sqrt{1- \rho_{1}^2(p,M)}.
 \end{eqnarray}
It is interesting to note that, regarded as conventional statistics, only
the estimator given in Eq.\,(\ref{bDTF}) shows a correction for a bias.

{\it Step}\,2) The {\sc ml} statistic of $H_{0}$ is given by
 \begin{equation}\label{H0Direct}
		{\cal H}^{\rm DTF} = \langle y \rangle_{2} -
 \beta\left(\sigma_{\zeta}^{\rm DTF}\right)^2,
 \end{equation}
and has a standard deviation equal to
\begin{equation}\label{DTFAccuracy}
\sigma_{{\cal H}^{\rm DTF}} = \frac{\sigma_{\zeta}^{\rm
DTF}}{\sqrt{N_{2}}}, \end{equation}
see Eq.\,(\ref{zetaDTF}).

The above statistics are no longer valid when selection effects on $p$, and as
well as
on $\mu$, are present. Nevertheless, they can easily be adapted by rewriting
the $p$-{\it
pdf} as $\tilde{f}_{p}(p) \propto \phi_{p}(p)f_{p}(p;p_{0},\sigma_{p})$, for
taking into
account $p$-selection effects. Let us emphasize that the correction in
Eq.\,(\ref{H0Direct}) is not of Malmquist type, in contrast with the one in
Eq.\,(\ref{bDTF}), see Section\,\ref{DistancesOfGalaxies}. Moreover, it must be
noted that
the magnitude of the bias does not depend on the limiting magnitude $m_{\rm
lim}$, and
of $\sigma_{p}$ (or equivalently $\sigma_{M}$).

\subsection{Comparison of estimators}\label{DiscussionModel}
It is important to understand that the Hubble constant $H_{0}$ has a similar
status among
these models, contrarily to parameters $a$ and $b$ which define the data
distribution on
the TF\,diagram. Indeed, it must be noted that the ITF model inherits all model
parameters
defined in the ITF$^\star$ model, simply because it is a particular case, where
the
functions $\phi_{m}(m)$, $\kappa(\mu)$ and $f_{M}(M)$ are specified. Hence, it
is clear
that $H_{0}$ keeps an identical status. On the other hand, because the {\rm
DTF} model
and the {\rm ITF} models describe the data distribution in a different way (see
Appendix\,\ref{Difference}), we might expect to obtain a different status.
However, let us
note that the {\it pd} given in Eq.\,(\ref{BasicThC}) is the projection of the
ITF$^\star$ {\it pd}, as well as the DTF one. Therefore, the model parameters
which are
in common (i.e., which are not cancelled by the projection), are identical
among these
models, which is the case of $H_{0}$.

Therefore, according to previous sections, which show that
Eq.\,(\ref{H0InverseC},\ref{H0InverseS},\ref{H0Inverse},\ref{H0Direct}) define
unbiased
$H_{0}$~statistic, the related estimates (for a given sample) correspond as a
matter of fact to
the same quantity, and the discrepancies (between these different estimates)
interpret as
statistical fluctuations which should vanish when the sample size increases.
With this in mind,
we investigate the nature of such discrepancies. These quantities can be
derived from the
following ones
 \begin{eqnarray}
\label{DeltaITFvITFS}
\Delta_{\rm II^\star} &=& {\cal H}^{\rm ITF}- {\cal H}^{\rm ITF^\star},\\
 \label{DeltaCvITFS}
 \Delta_{\rm CI^\star} &=& {\cal H}^{\rm C}-{\cal H}^{\rm ITF^\star},\\
\label{DeltaDTFvITFS}
\Delta_{\rm DI^\star} &=& {\cal H}^{\rm DTF}- {\cal H}^{\rm ITF^\star},
 \end{eqnarray}
where the ITF$^\star$ is chosen as a reference estimate. According to
Eq.\,(\ref{DeltaITFvITFS},\ref{DeltaCvITFS}), the difference between the
statistics given
in Eq.\,(\ref{H0InverseS},\ref{H0Inverse}) reads
 \begin{equation}\label{H0CompIS}
		\Delta_{\rm II^\star} =
\frac{\gamma^{2}}{1+\gamma^{2}}\Delta_{\rm CI^\star}
 \end{equation}
where $\gamma=\gamma^{\rm ITF}$, see Eq.\,(\ref{SigmaZetaInequal}). Thus, the
smaller the
ratio $\gamma^{\rm ITF}$, the smaller the discrepancy $\Delta_{\rm II^\star}$.
Let us
elucidate this particular property, which shows clearly the gain of knowledge
on $H_{0}$
when the TF relation is used. It is important to note that the estimates ${\cal
H}^{\rm
C}$ and ${\cal H}^{\rm ITF^\star}$, given in
Eq.\,(\ref{H0InverseC},\ref{H0CompIS}), are
independent, i.e., they involve two different types of information. Indeed, the
${\cal
H}^{\rm C}$ is based only on characteristics related to the luminosity
distribution
function of sources (or equivalently, on the $p$-distribution), while the
${\cal H}^{\rm
ITF^\star}$ takes into account only the TF\,relation (i.e., the
$\zeta$-distribution).
When both features are used, we obtain a more accurate estimate ${\cal H}^{\rm
ITF}$,
which lies between ${\cal H}^{\rm C}$ and ${\cal H}^{\rm ITF^\star}$. Actually,
it lies
much more close to ${\cal H}^{\rm ITF^\star}$, accordingly to
Eq.\,(\ref{SigmaZetaInequal}), which shows that this estimator is less
sensitive to
hypotheses on the luminosity distribution function of sources, which translates
the
robustness of the estimator ${\cal H}^{\rm ITF^\star}$.

In order to calculate $\Delta_{\rm DI^\star}$, let us compare the statistics
(\ref{aITF}-\ref{zetaITF}) and (\ref{aDTF}-\ref{zetaDTF}). After little
algebra, we
obtain
 \begin{eqnarray}
\label{aComp} a^{\rm DTF} &=& \rho_{1}^2(p,M) \times a^{\rm ITF},\\
\label{bComp} b^{\rm DTF} &=& b^{\rm ITF}+\left(1-
 \rho_{1}^2(p,M)\right)\left(a^{\rm ITF}\langle p \rangle_{1} + \beta
(\Sigma_{1}(M))^2\right), \\
\label{SigmaZetaComp} \sigma_{\zeta}^{\rm DTF} &=& \left|\rho_{1}\right|\times
 \sigma_{\zeta}^{\rm ITF}.
 \end{eqnarray}
Hence, according to
Eq.\,(\ref{zetaITF},\ref{H0InverseS},\ref{H0Direct},\ref{DeltaDTFvITFS}), it
follows\footnote{Let us remind that the average $\langle y \rangle_{2}$ is a
model-dependent quantity.}
\begin{equation}\label{H0CompDS}
		\Delta_{\rm DI^\star} = \frac{\rho_{1}}{\left|\rho_{1}\right|}\:
 {\cal C}_{p}\:\frac{\gamma}{\sqrt{1+\gamma^{2}}}\:\sigma_{\zeta}^{\rm ITF}
 \end{equation}
 where $\gamma=\gamma^{\rm ITF}$, see Eq.\,(\ref{SigmaZetaInequal}),
$\rho_{1}=\rho_{1}(p,M)$ and
 \begin{equation}\label{P12}
{\cal C}_{p} = \frac{\langle p \rangle_{1} - \langle p
\rangle_{2}}{\Sigma_{1}(p)}.
 \end{equation}
Thus $\Delta_{\rm DI^\star}$ depends essentially on two independent
characteristics. The
first one is the discrepancy of sample averages of $p$~values between the
calibration
sample and the one used to determine $H_{0}$. The second one is the accuracy of
the
TF~relation, and similarly as above, Eq.\,(\ref{H0CompDS}) shows that the
smaller the
ratio $\gamma^{\rm ITF}$ the smaller the discrepancy. Let us emphasize that
such a
feature can also be interpreted in terms of $M$--$p$ correlation, since we have
$\gamma^{2}/(1+\gamma^{2})\approx1-\rho_{1}(p,M)^{2}$, see
Eq.\,(\ref{SigmaZetaInequal},\ref{zetaITF}). Thus, the higher the value of
$\left|\rho_{1}(p,M)\right|$ the smaller the discrepancies given in
Eq.\,(\ref{H0CompIS},\ref{H0CompDS}). The hypothetical case
$\left|\rho_{1}(p,M)\right| =
1$ is a singular situation, where the data distribution on the TF~diagram
coincides with
the straight line ($\Delta_{\rm TF}$) defined by Eq.\,(\ref{Rough}), which
makes the
ITF$^{\star}$, the ITF and DTF approaches identical.

Now let us calculate the expected orders of magnitude of discrepancies given in
Eq.\,(\ref{DeltaITFvITFS}-\ref{DeltaDTFvITFS}), and their dependence on the
sample size.
Note that the statistics defined in Eq.\,(\ref{H0InverseC},\ref{H0InverseS}),
considered
as random variables (see Sec.\,\ref{DeterminationITF}), are independent. Hence,
$\Delta_{\rm CI^\star}$, resp. $\Delta_{\rm II^\star}$, both have a vanishing
expected
value, with a standard deviation
 \begin{equation}\label{StDevCS}
\sigma_{\Delta_{\rm CI^\star}} \approx
\sqrt{1+\gamma^2}\frac{\sigma_{M}}{\sqrt{N_{2}}},
 \end{equation}
resp.
 \begin{equation}\label{StDevIS}
\sigma_{\Delta_{\rm II^\star}} \approx
\frac{\gamma}{\sqrt{1+\gamma^2}}\frac{\sigma_{\zeta}^{\rm ITF}}{\sqrt{N_{2}}}.
 \end{equation}
Therefore, the difference between the ITF and the ITF$^{\star}$ estimates is
not
systematic, and thus there is no bias. Moreover these estimates coincide as the
sample size
$N_{2}$ increases. Similarly, since $\langle p \rangle_{1}$, resp. $\langle p
\rangle_{2}$,
is a statistic providing the mean $p_{0}$ with a standard deviation of
$\sigma_{p}/\sqrt{N_{1}}$, resp. $\sigma_{p}/\sqrt{N_{2}}$, and that {\it
Step},1 and {\it
Step}\,2 are independent, then ${\cal C}_{p}$ is a random variable of vanishing
mean with
standard deviation $\approx \sqrt{1/N_{1}+1/N_{2}}$. Hence, $\Delta_{\rm
DI^\star}$, has a
vanishing expected value, with a standard deviation
 \begin{equation}\label{StDevDS}
\sigma_{\Delta_{\rm DI^\star}}=
\frac{\gamma}{\sqrt{1+\gamma^2}}\:\sigma_{\zeta}^{\rm
ITF}\sqrt{1/N_{1}+1/N_{2}}.
 \end{equation}
Thus there is no bias between the DTF and the ITF$^{\star}$ estimates, while
the
estimates of model parameters $a$ and $b$ are different, see
Eq.\,(\ref{aComp}-\ref{SigmaZetaComp}). Nevertheless, it must be noted that
these estimates
coincide only when both sample sizes $N_{1}$ and $N_{2}$ increase, which
emphasizes the
importance of the calibration of the TF\,relation.

Let us remind that the ITF and the DTF models belong to a single class of
models defined
by Eq.\,(\ref{TFClass}). Intermediate models can be obtained by means of a
rotation
parameter which makes a link between the ITF and the DTF models. Thus, we
understand that
simple arguments (of linearity) indicate that the $H_{0}$ statistics related to
these
models provide asymptotically identical estimates, Triay
et\,al.\,(\cite{TriEtal93b}). More
generally, we might ask whether this is still true for models describing the TF
diagram in a more
complex way. The element of answer comes by noting that we have $y = \zeta +
{\cal H}$,
see Eq.\,(\ref{Rough},\ref{TFrelation},\ref{variableY}), and that $P_{\rm
th}(y) = {\cal
H}$ independently of hypothesis ($h_{3}$), which forces the random variable
$\zeta$ to
have a vanishing mean value. Thus, we can claim that if this condition is
complete then
the way of describing the data does not influence the determination of $H_{0}$.

On account of these results, let us ask whether the accuracy of estimates may
be used as
criterion for having a preference for a particular model. According to
Eq.\,(\ref{SigmaZetaInequal},\ref{ITFSAccuracy},\ref{ITFAccuracy},
\ref{DTFAccuracy},\ref{SigmaZetaComp}),
it turns out that we obtain
 \begin{equation}\label{AccuracyCompITFDTF}
		\sigma_{{\cal H}^{\rm DTF}} \approx \sigma_{{\cal H}^{\rm ITF}},
 \end{equation}
while the ${\rm ITF^\star}$ estimate is less accurate, see
Eq.\,(\ref{ITFAccuracy}). Thus, we have
similar precisions in estimating $H_{0}$ by using indifferently either the {\rm
DTF} or the {\rm ITF}
approaches. Work is in progress for checking whether an intermediate
statistical model which
describes the TF relation might provide higher accuracy on the determination of
$H_{0}$, Triay
et\,al.\,(\cite{TriEtal93b}).

According to Eq.\,(\ref{H0CompDS}), it is interesting to note that if the
following
equality
 \begin{equation}\label{CpCritera}
{\cal C}_{p} = 0,
 \end{equation}
herein called ``${\cal C}_{p}$-{\it criteria}'', is fulfilled for a given
sample, then the
{\it algebraic} expressions of the ITF$^{\star}$ and the DTF estimators of
$H_{0}$ become
identical. On the other hand, if the ``${\cal C}_{p}$-{\it criteria}'' is not
verified then
nothing prevents us to resample (to remove data according to rules allowed by
the working
hypotheses) on this purpose (although it reduces the sample size, and thus
diminish the
information). Namely, the faintest objects can be removed until
Eq.\,(\ref{CpCritera}) is
complete, which preserves the selection rule based on the magnitude selection
effect (i.e.,
the hypothesis $h_{1}$), but with a brighter limiting magnitude.

\subsection{Applications}\label{Simulation}
In order to have a visual support for our theoretical approach and to
investigate the
influence of calibration errors in the determination of $H_{0}$, we perform
$N_{\rm S} =
1\,000$ simulations. We generate two sorts of samples~: the $\{ m_{k}, \mu_{k},
p_{k}\}_{k=1,N_{1}}$, which is involved in the calibration step, and the
$\{m_{k}, \eta_{k},
p_{k}\}_{k=1,N_{2}}$, which is involved in the determination of the Hubble
constant $H_0$,
this one contains $N_{2}=100$ objects. For each sample, we have three
independent data
sets~: the absolute magnitudes $\{M_{k}\}$, the intrinsic TF dispersions
$\{\zeta_{k}\}$,
according to working hypotheses ($h_{3}$,$h_{4}$), see
Eq.\,(\ref{TFClass},\ref{MGaussian}),
and the distance moduli $\{\mu_{k}\}$, according to working hypotheses
($h_{1}$,$h_{2}$),
see Eq.\,(\ref{uniform}); the $\{p_{k}\}$ are derived from
Eq.\,(\ref{Rough},\ref{TFrelation}). The apparent magnitudes $\{m_{k}\}$ are
given by
Eq.\,(\ref{ApparentMagnitude}), and the cosmological velocity moduli
$\{\eta_{k}\}$ are
calculated according to Eq.\,(\ref{HubbleL}), with a Hubble constant given by
\begin{equation}
\label{H0Simul} H_{0}^{\rm S}=100\:{\rm km\,s^{-1}/Mpc}.
\end{equation}
The characteristics of samples are the following~:
 \begin{itemize} \item completeness of samples, up to
a limiting magnitude of
 \begin{equation} \label{LimitingMagnitude}
 m_{\rm lim} = 12;
\end{equation}
\item a uniform spatial distribution of sources;
\item a Gaussian luminosity distribution function, defined by
\begin{equation} \label{SimulMagnitude} M_{0} = -19 \quad ,\quad
\sigma_{M} = 1.5;
 \end{equation}
\item we assume (a priori) the ITF model, so that the TF diagram shows a normal
$\zeta$-dispersion at constant $M$ defined by
 \begin{equation} \label{SimulZeta}
\sigma_{\zeta} = 0.5,
\end{equation}
and with the following calibration parameters
\begin{equation}\label{TFparameters}
a = -6 \quad {\rm and} \quad b = -7.
\end{equation}
\end{itemize}

A first set of simulations is performed in order to investigate the
$H_{0}$\,statistics
regardless of calibration errors. In this case, a unique calibration sample is
used, the sample size of $N_{1}^{(1)} = 8\,000$ galaxies is large enough so
that the
estimates of model parameters $a$, $b$ and $\sigma_{\zeta}$, as given by
Eq.\,(\ref{aITF}-\ref{zetaITF},\ref{aDTF}-\ref{zetaDTF}), are expected to be
free of
statistical fluctuations.
\begin{table}[htbp]
\caption{Comparison of estimators. The estimates of parameters $a$, $b$ and
$\sigma_{\zeta}$
are obtained from a sample of $N_{1}^{(1)} = 8\,000$ galaxies (which gives
values
which are free of statistical fluctuations). The related standard deviations
are
obtained from 30 trials of such samples. The estimate of $H_{0}$ corresponds to
the mean
value on $N_{\rm S} = 1\,000$ trials with samples of $N_{2}=100$ objects.}
 \begin{flushleft}
\begin{tabular}{|c|ccc|}
\hline
Parameter & ITF$^\star$ & ITF & DTF \\
\hline
$a$ & \multicolumn{2}{c}{$-5.99\pm0.02$} & $-5.39\pm0.02$ \\
$b$ & \multicolumn{2}{c}{$-7.02\pm0.04$} & $-8.22\pm0.05$ \\
$\sigma_{\zeta}$ & \multicolumn{2}{c}{$0.500\pm0.004$} & $0.475\pm0.003$ \\
$H_{0}-H_{0}^{\rm S}$ & $-0.2\pm2.4$ & $-0.2\pm2.2$ & $-0.2\pm2.2$ \\
\hline
\end{tabular}
\end{flushleft}
\label{Comparaison}
\end{table}
The results are given in Table\,\ref{Comparaison}. It is reassuring to note
that the ITF
estimates of model parameters $a$, $b$ and $\sigma_{\zeta}$, correspond as a
matter of fact
to values used to generate the random samples, see Eq.\,(\ref{TFparameters}),
while it is
not the case for the DTF estimates. These estimates are used to determine
$H_{0}$, according
to Eq.\,(\ref{H0InverseS},\ref{H0Inverse},\ref{H0Direct}), on $N_{\rm S} =
1\,000$ samples of
$N_{2}=100$ objects. The average of these estimates and their related accuracy
($1\sigma$)
are given in Table\,\ref{Comparaison}. In agreement with the theory, we can
note that these
statistics give back the value $H_{0}^{\rm S}$, used for the simulations, which
shows that
they are not biased. Moreover, we can note that the related accuracies are in
agreement with
the expected value obtained from
Eq.\,(\ref{ITFSAccuracy},\ref{ITFAccuracy},\ref{DTFAccuracy}), where we use
$\sigma_{\cal
H}\approx\Sigma\left({\cal H}\right)$.
\begin{figure}[htbp]
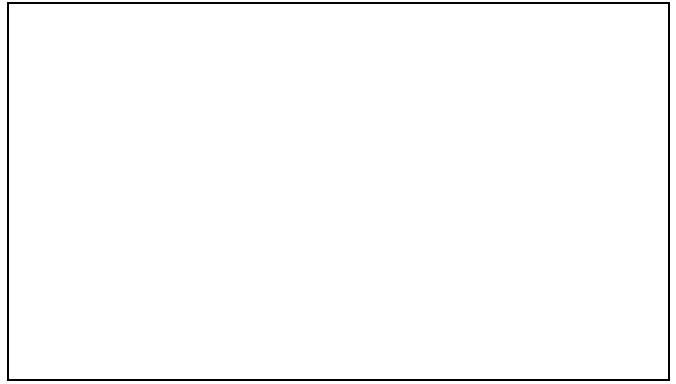

 \picplace{5cm}
 \caption{Comparison between the ITF, ITF$^\star$ and the DTF
estimates, without taking into account calibration errors.
The $x$-coordinate of symbols correspond to ITF estimates. The
$y$-coordinate of symbols ``$+$'' correspond to DTF estimates, while the
$y$-coordinate of symbols ``$\bigcirc$'' correspond to ITF$^\star$ estimates.}
 \label{ISDvI}
 \end{figure}
Figure\,\ref{ISDvI} shows 100 $H_{0}$-estimates from these different
approaches. The
symbols ``$+$'' correspond to the DTF versus the ITF estimate, and the symbols
``$\bigcirc$'' correspond to the ITF$^\star$ versus the ITF estimate. The
evident
distribution along the diagonal shows that these approaches are as a matter of
fact
equivalent. Moreover, we can speculate that it is advantageous to prefer the
ITF$^\star$
approach, which is more robust, since there is no real gain of accuracy by
choosing the
other ones. It is interesting to note that the distribution of symbols
``$\bigcirc$'' is
more scattered about the diagonal than the symbols ``$+$''. The differences of
accuracy
between these estimates don't suffice to account for such a gap (which can be
estimated
of the order of 0.01). Therefore, this means that the ITF$^{\star}$-ITF
estimates are less
correlated than the DTF-ITF ones, while one might expect the opposite. Indeed,
let us
remind that the ITF model is defined from the ITF$^{\star}$ model by specifying
the
functions $\phi_{m}$, $\kappa$ and $f_{M}$, and thus it is simply a particular
case of the
ITF$^{\star}$ model, while it is different from the DTF model. The reason lies
in the
amount of information which is used by these estimators. Indeed, the
ITF$^{\star}$
estimator uses less working hypotheses than the ITF and the DTF ones, which
makes it more
slacker.
\begin{table}[htbp] \caption{Effects due to calibration errors. These estimates
correspond to mean values calculated on $N_{\rm S} = 1\,000$ trials. The
parameters $a$,
$b$ and $\sigma_{\zeta}$ are obtained from samples of $N_{1}^{(2)} = 30$
galaxies, while
$H_{0}$ is determined from samples of $N_{2}=100$ objects.}
 \begin{flushleft}
\begin{tabular}{|c|ccc|}
\hline
Parameter & ITF$^\star$ & ITF & DTF \\
\hline
$a$ & \multicolumn{2}{c}{$-6.06\pm0.40$} & $-5.40\pm0.32$ \\
$b$ & \multicolumn{2}{c}{$-6.84\pm1.02$} & $-8.21\pm0.85$ \\
$\sigma_{\zeta}$ & \multicolumn{2}{c}{$0.498\pm0.075$} & $0.469\pm0.064$ \\
$H_{0}-H_{0}^{\rm S}$ & $-0.2\pm4.4$ & $-0.2\pm4.0$ & $-0.1\pm4.0$ \\
\hline
\end{tabular}
\end{flushleft}
\label{CalibErr}
\end{table}

In general, the calibration of the TF\,relation is performed only on few tens
of galaxies, which
makes the estimates of model parameters $a$, $b$ and $\sigma_{\zeta}$ much less
precise. Hence,
the determination of $H_{0}$ undergoes the related statistical fluctuations,
herein called {\it
calibration errors}. So such effects are investigated by using simulated
calibration samples,
with a more realistic sample size of $N_{1} = 30$ galaxies, and by determining
$H_{0}$ on
samples of $N_{2}=100$ objects. The statistical analysis is performed on
$N_{\rm S} = 1\,000$
trials. The results are shown in Table\,\ref{CalibErr}, which gives the
averages of
parameters estimates, and their related accuracy ($1\sigma$). The comparison
with
Table\,\ref{Comparaison} shows that the estimates of model parameters $a$, $b$
and
$\sigma_{\zeta}$ are similar, and that the estimation of $H_{0}$ is not biased
by
calibration errors, while it is obviously less accurate.
\begin{figure}[htbp]
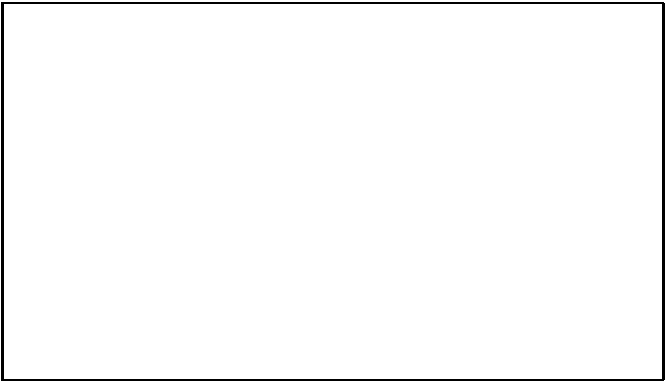

 \picplace{5cm}
 \caption{Comparison between the ITF, ITF$^\star$ and the DTF\,estimates, by
taking into account calibration errors. The same caption as in
Fig.\,1.}
 \label{ISDvIwCE}
 \end{figure}
Figure\,\ref{ISDvIwCE} shows more elongated distributions than those in
Fig.\,\ref{ISDvI}, but still about the diagonal. At first glance, the main
result is that
the above conclusions are still valid when calibration errors are taken into
account
(actually, it is easy to note that these approaches are even much more
equivalent, since
the scatter of symbols ``$+$'' about the diagonal is now comparable to the one
of
symbols ``$\bigcirc$'').

\section{About measurement errors}\label{Errors}
It is clear that Malmquist bias is present only when a part of the (luminosity)
{\it pdf}
is not observed. Since this is independent of errors distribution, we
understand that
measurement errors do not produce such a bias, see ({\it Def.}6).
Notwithstanding, in
order to answer the question of whether these effects introduce another type of
bias, we
must use a precise mathematical framework for avoiding misunderstandings.

\subsection{The statistical model}\label{ErrorsModel}
Let $\left\{\epsilon_{m},\epsilon_{\mu},\epsilon_{p},\epsilon_{\eta}\right\}$
denote the measurement errors. We use the symbol hat ``$\hat{\;}$'' to
distinguish the ({\it measurable}) variables (i.e. the ones which are affected
by
measurement errors), from formal ones which are given by
 \begin{eqnarray}
 m &=& \hat{m} - \epsilon_{m}, \label{errorsm} \\
 \mu &=& \hat{\mu} - \epsilon_{\mu}, \label{errorsmu} \\
 p &=& \hat{p} - \epsilon_{p}, \label{errorsp} \\
 \eta &=& \hat{\eta} - \epsilon_{\eta}. \label{errorseta}
 \end{eqnarray}
If these errors are independent Normal random variables, they are distributed
according to the following the {\it pd}
\begin{equation}\label{BasicError0}
	dP_{\epsilon}^{(s)} =
\prod_{\lambda \in \Lambda^{(s)}}g_{\rm
 G}(\epsilon_{\lambda};0,\sigma_{\epsilon_{\lambda}}^{(s)})d\epsilon_{\lambda},
 \end{equation}
where the index $\lambda$ takes character values among the set $\Lambda^{(1)} =
\left\{m,\mu,p\right\}$ or $\Lambda^{(2)} = \left\{m,\eta,p\right\}$, depending
on {\it Step}\,1 or {\it Step}\,2. Since a magnitude limited sample can be
selected from a catalog, where the data are already affected by measurement
errors, we easily understand that the selection function must be written in
term
of measurable observables. Therefore, according to Eq\,(\ref{BasicObs}), the
{\it pd} which describes both the observables and the random errors is given by
 \begin{equation}\label{BasicError1}
	d\hat{P}_{\rm obs}^{(s)} =
\frac{\hat{\phi}_{m}}{P_{\rm th}\left(P_{\epsilon}^{(s)}\left(\hat{\phi}_{m}
\right)\right)} ~dP_{\rm th}\times dP_{\epsilon}^{(s)},\quad
\left(s=1,2\right),
 \end{equation}
where
 \begin{equation}\label{SelectEffectErr}
\hat{\phi}_{m}(m) = \phi_{m}(\hat{m}).
 \end{equation}
It is clear that since the errors $\epsilon_{\lambda}$, see
Eq.\,(\ref{errorsm}-\ref{errorseta}), cannot be disentangled from intrinsic
scatter, the {\sc ml} technique is not feasible for obtaining genuine
statistics. However, we can overcome this obstacle by substituting the suitable
corrections by their expected values, which can be calculated according
to the {\it pd} given in Eq.\,(\ref{BasicError1}). For convenience in writing,
we
use the following dimensionless quantities
 \begin{eqnarray}
 \delta_{p} &=&
\frac{\left(\sigma_{\epsilon_{p}}^{(1)}\right)^{2}}{(\Sigma_{1}(\hat{p}))^{2}},
 \label{DeltaErrorsp} \\
 \delta_{M} &=&
\frac{\left(\sigma_{\epsilon_{m}}^{(1)}\right)^{2}
+\left(\sigma_{\epsilon_{\mu}}^{(1)}\right)^{2}}{(\Sigma_{1}(\hat{M}))^{2}},
\label{DeltaErrorsM}
 \end{eqnarray}
see Eq.\,(\ref{ApparentMagnitude}), and $\widehat{{\cal C}}_{p}$, see
Eq.\,(\ref{P12}). It turns
out that the spatial distribution of sources must be specified (i.e.,
$\kappa(\mu)$) a priori in
order to perform such calculations, see Appendix\,\ref{ErrorsCorrection}. If we
assume that it
is uniform ($h_{2}$), see Eq.\,(\ref{uniform}), then the normalization term is
given by
\begin{equation}\label{NormaliErrors}
P_{\rm th}\left(P_{\epsilon}^{(s)}\left(\hat{\phi}_{m}\right)\right) = P_{\rm
th}\left(\phi_{m}\right) \exp{ \left(\frac{1}{2}\left(\beta
\sigma_{\epsilon_{m}}^{(s)} \right)^{2}\right)},
 \end{equation}
where $P_{\rm th}\left(\phi_{m}\right)$ is the normalization term when
measurement
errors are not taken into account. The effects due to measurement errors lie
only in the
extra term, which turns out to be independent on model parameters, and thus
which ensures
the absence of Malmquist bias on estimating these parameters. Nevertheless,
there are
biases of different nature in the $H_{0}$\,estimates, which are given in
Eq.\,(\ref{H0InverseC},\ref{H0InverseS},\ref{H0Direct}), since the calculation
provides us
with
\begin{eqnarray}
 \label{H0ITFSErrors}
		{\cal H}^{\rm ITF^{\star}} &=& \widehat{{\cal H}}^{\rm
ITF^{\star}}
 + \widehat{{\cal C}}_{p}\delta_{M}\hat{a}^{\rm ITF}\Sigma_{1}(\hat{p})
\nonumber \\
 &+& \beta\left(\left(\sigma_{\epsilon_{m}}^{(1)}\right)^{2} -
 \left(\sigma_{\epsilon_{m}}^{(2)}\right)^{2}\right),\\
 \label{H0ITFCErrors}
		{\cal H}^{\rm C} &=& \widehat{{\cal H}}^{\rm C}
 + \beta\left(\sigma_{\epsilon_{m}}^{(2)}\right)^{2}, \\
 \label{H0DTFErrors}
		{\cal H}^{\rm DTF} &=& \widehat{{\cal H}}^{\rm DTF} +
\widehat{{\cal C}}_{p}\frac{\delta_{p}}{1-\delta_{p}}
 \hat{a}^{\rm DTF}\Sigma_{1}(\hat{p}) \nonumber \\
 &+& \beta\left(\left(\sigma_{\epsilon_{m}}^{(1)}\right)^{2}
- \left(\sigma_{\epsilon_{m}}^{(2)}\right)^{2}\right).
 \end{eqnarray}
The bias free ITF statistics is obtained by substituting in
Eq.(\ref{H0Inverse}) the terms
${\cal H}^{\rm ITF^{\star}}$ and ${\cal H}^{\rm C}$, as given by
Eq.(\ref{H0ITFSErrors},\ref{H0ITFCErrors}). Therefore, we see that the
statistics given in
Eq.\,(\ref{H0InverseC},\ref{H0InverseS},\ref{H0Inverse},\ref{H0Direct}) can be
restored as long
as the standard deviations of errors are known, i.e., $\sigma_{\epsilon_{m}}$
and
$\sigma_{\epsilon_{\mu}}$ in the case of the ITF model, and
$\sigma_{\epsilon_{p}}$ in the case
of the DTF model\footnote{Note that these criteria can motivate the choice of
the model to be
used.}. However, it is clear that these corrections should be tiny quantities,
since
$\delta_{p}\ll1$ and $\delta_{M}\ll1$, unless the information is buried into
noise. Moreover, it
is interesting to note that they are of different nature, the first one depends
on TF
characteristics and can be removed by using the ${\cal C}_{p}$-{\it criteria},
see
Eq.\,(\ref{CpCritera}), while the other one does not.

\subsection{Applications}
\begin{figure}[htbp]
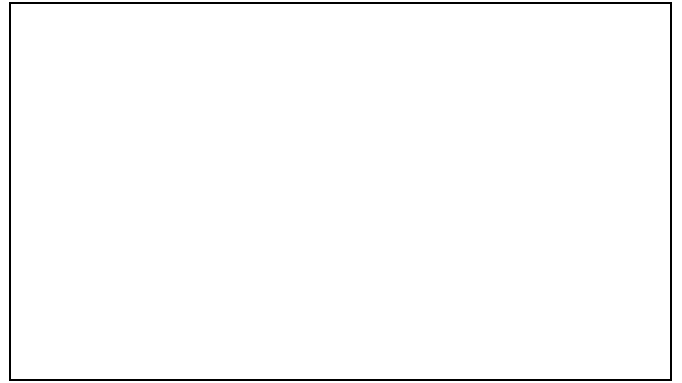

 \picplace{5cm}
 \caption{Comparison between the ITF, ITF$^\star$ and the DTF estimates. The
calibration
error are taken into account. The same caption as in Fig.\,1.}
 \label{ISDvIwME}
 \end{figure}

\begin{figure}[htbp]
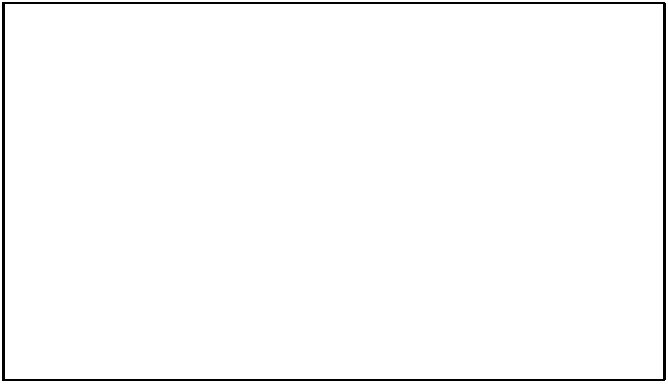

 \picplace{5cm}
 \caption{Comparison between the ITF, ITF$^\star$ and the DTF estimates. The
calibration
errors are taken into account. The same caption as in Fig.\,1.}
 \label{ISDvIwMEwCE}
 \end{figure}

Similarly to Section ~\ref{Simulation}, we perform simulations in order to
enlighten on above
results and to investigate the effects of measurement errors on the accuracy of
estimates.
According to working hypotheses, we use simulated samples with characteristics
given by
Eq.\,(\ref{H0Simul}-\ref{TFparameters}), where the observables are perturbed by
normal random
errors. In practice, according to Gouguenheim\,(\cite{Gou93}), the observables
are measured
within the following accuracies~:
\begin{itemize}
\item the line width (which gives $p$) is measured within
20\,km/s;
\item the recession of galaxies is given within 15\,km/s;
\item for calibrators ({\it Step}\,1), the apparent magnitudes are measured
within an accuracy of 0.05 mag., while for {\it Step}\,2, the accuracy
depends on magnitude, it is of order of 0.1 mag. for $m \leq 13$, of 0.15
mag. for $13 < m \leq 14$, and of 0.2 mag. for $m > 14$.
\end{itemize}
These above uncertainties can be interpreted as the 2-3 errors standard
deviations. We can use a good compromise on the magnitude of errors for
avoiding their dependence on the magnitude of the related variable by assuming
the following characteristics~:
 \begin{equation}\label{SimulError1}
 \sigma_{\epsilon_{m}}^{(1)} = 0.05,\quad
 \sigma_{\epsilon_{p}}^{(1)} = 0.025,\quad
 \sigma_{\epsilon_{\mu}}^{(1)} = 0.15,
 \end{equation}
and for samples used to determine $H_{0}$, according to
Fouqu\'e\,(\cite{Fou93}), we choose
 \begin{equation}\label{SimulError2}
 \sigma_{\epsilon_{m}}^{(2)} = 0.15,
 \quad\sigma_{\epsilon_{p}}^{(2)} = 0.025,
 \quad \sigma_{\epsilon_{\eta}}^{(2)} = 0.0.
 \end{equation}
The effect on $H_{0}$\,statistics due to measurement errors, but regardless of
calibration
errors, is investigated by using a unique calibration sample with $N_{1}^{(1)}
= 8\,000$
galaxies, and a sample of $N_{2}=100$ objects for the $H_{0}$ determination.
Theses samples are
generated according to Eq.\,(\ref{H0Simul}-\ref{TFparameters}), and both are
perturbed by normal
random errors with characteristics defined by
Eq.\,(\ref{SimulError1},\ref{SimulError2}). The
statistical analysis is performed on $N_{\rm S} = 1\,000$ trials. The
statistics of model
parameters $a$, $b$ and $\sigma_{\zeta}$, corrected for the bias due to
measurement errors are
defined in Eq.\,(\ref{aITFErrors}-\ref{zetaDTFErrors}). The related results are
given in
Table\,\ref{ErrorsTab}, which shows the averages of model parameters estimates
and their related
accuracies (1 $\sigma$), and the magnitude of the correction terms which are
present in the
$H_{0}$\,statistics, given in Eq.\,(\ref{H0ITFSErrors}-\ref{H0DTFErrors}). We
can note that these
corrections are effective since the mean value of $H_{0}$\,estimates gives back
the value
$H_{0}^{\rm S}$. However, it is clear that this is a minor quantity compared to
the
$H_{0}$\,standard deviation.
\begin{table}
\caption{Measurement errors, without calibration errors. The results are based
on $N_{\rm S} =
1\,000$ trials. The parameters $a$, $b$ and $\sigma_{\zeta}$, obtained from
samples of $N_{1}^{(1)} =
8\,000$ galaxies, while $H_{0}$ is measured from samples of $N_{2}=100$
objects. The correction term
$\Delta_{\epsilon}H_{0}$ is writen in $H_{0}$ unit.}
\begin{flushleft}
\begin{tabular}{|l|ccc|} \hline Parameter & ITF$^\star$ & ITF & DTF \\ \hline
$a$ &
\multicolumn{2}{c}{$-6.00\pm0.03$} & $-5.40\pm0.02$ \\
 $b$ & \multicolumn{2}{c}{$-6.99\pm0.07$} & $-8.20\pm0.06$\\
$\sigma_{\zeta}$ & \multicolumn{2}{c}{$0.501\pm0.005$} & $0.475\pm0.004$ \\
 $H_{0}-H_{0}^{\rm S}$ & $-1.2\pm2.5$ & $-1.2\pm2.4$ & $-1.2\pm2.4$\\
 $\Delta_{\epsilon}H_{0}$ & $-1.27\pm0.08$ & $-1.27\pm0.07$ & $-1.27\pm1.79$ \\
 \hline
\end{tabular}
\end{flushleft}
\label{ErrorsTab}
\end{table}
Figure\,\ref{ISDvIwME} shows the same diagram as in Fig.\,\ref{ISDvI}. The
comparison between
these figures indicates that the measurement errors do not perturb the
correlation between the
ITF$^\star$, the ITF and the DTF methods.
A similar analysis is performed by taking into account simultaneously
calibration errors. The
method of proceeding is identical to Section\,\ref{Simulation} above. The
results are given in
Table\,\ref{ErrorsTabWCE}, and in Fig.\,\ref{ISDvIwMEwCE}. The main effect of
calibration errors
is to increase the standard deviation of both the correction term and $H_{0}$,
which does not
change our previous conclusions, while the ITF estimate seems to be 10 percent
more accurate.
\begin{table}
\caption{Measurement errors, with calibration errors. The same caption as in
Table\,3.}
\begin{flushleft}\begin{tabular}{|l|ccc|}
\hline
Parameter & ITF$^\star$ & ITF & DTF \\
\hline
$a$ & \multicolumn{2}{c}{$-6.03\pm0.42$} & $-5.43\pm0.35$ \\
$b$ & \multicolumn{2}{c}{$-6.90\pm1.06$} & $-8.12\pm0.90$ \\
$\sigma_{\zeta}$ & \multicolumn{2}{c}{$0.482\pm0.084$} & $0.457\pm0.073$ \\
$H_{0}-H_{0}^{\rm S}$ & $-0.6\pm5.2$ & $-0.6\pm4.8$ & $-0.8\pm5.0$\\
$\Delta_{\epsilon}H_{0}$ & $-1.29\pm0.13$ & $-1.29\pm0.11$ & $1.25\pm1.77$ \\
\hline
\end{tabular}
\end{flushleft}
\label{ErrorsTabWCE}
\end{table}

\section{The distances of galaxies and $H_{0}$}\label{DistancesOfGalaxies}
Within the same framework, let us investigate the problem of finding a reliable
determination of
distances of galaxies, by using the TF relation. This still involves a first
step of calibration
of the TF relation, and thus we refer to above results. On the other hand, the
next step does
not involve a second sample, but only a unique galaxy, whose the related data
are denoted by
$m_k$ and $p_k$. At first glance, according to
Eq.\,(\ref{ApparentMagnitude},\ref{Rough}), the
distance modulus of a galaxy of known apparent magnitude $m_{k}$ and line width
distance
estimator $p_{k}$, can be estimated by the following statistic
 \begin{equation}\label{AHDistanceEstimate}
\tilde{\mu}_{k} = m_{k} - (a.p_{k} + b).
 \end{equation}
We easily understand that, because $a$ and $b$ are (as a matter of fact) model
dependent
parameters, similarly to the above approach, the question of whether this (ad
hoc)
estimate is biased takes its answer only within a given model. Thus we have to
define a
{\it pd} which describes the distribution of distance modulus $\mu$ provided
$m=m_{k}$
and $p=p_{k}$, that we denote $dP_{\mu}^{(k)} =
f_{\mu}(\mu;\mu_{0}^{(k)},\sigma_{\mu}^{(k)}) d\mu$. According to
Eq.\,(\ref{BasicObs}), by
using conditional probabilities, we have
 \begin{equation}\label{BasicDist}
dP_{\mu}^{(k)} = \frac{\delta(m-m_{k})\delta(p-p_{k})}{P_{\rm
obs}\left(\delta(m-m_{k})\delta(p-p_{k})\right)}\,dP_{\rm obs},
 \end{equation}
and thus
 \begin{equation}\label{BasicDistFunc}
f_{\mu}(\mu;\mu_{0}^{(k)},\sigma_{\mu}^{(k)}) =
\frac{F(m_{k}-\mu,p_{k})~\kappa(\mu)}{\int
F(m_{k}-\mu,p_{k})~\kappa(\mu)d\mu}.
 \end{equation}
Since no selection function intervenes in this equation, we deduce that no
Malmquist bias
is present. Moreover, the expected value of the distance modulus, which is
given by
 \begin{equation}\label{RoughDistanceEstimate}
\mu_{0}^{(k)} = \int \mu dP_{\mu}^{(k)},
 \end{equation}
see Eq.\,(\ref{BasicDist}), provides us with the most likelihood statistic. It
is
clear that such an estimate depends on the form of the TF{\it pdf}, which reads
\begin{eqnarray}\label{TFpdf}
F(m_{k}-\mu,p_{k}) &=& g_{\rm G}(\mu;\tilde{\mu}_{k},\sigma_{\zeta})
\nonumber\\
&\times&\left\{
\begin{tabular}{ll}
$af_{M}(m_{k} - \mu;M_{0},\sigma_{M})$ & (ITF)
\\ $f_{p}(p_{k};p_{0},\sigma_{p})$ & (DTF)
\end{tabular}
 \right.
\end{eqnarray}
see
Eq.\,(\ref{InverseTFDensity},\ref{DirectTFDensity},
\ref{RoughDistanceEstimate}).
Hence, the {\it pd} given in
Eq.\,(\ref{BasicDist}), transforms
 \begin{eqnarray}\label{BasicDist1}
dP_{\mu}^{(k)} &\propto& g_{\rm G}(\mu;\tilde{\mu}_{k},\sigma_{\zeta})
\kappa(\mu)d\mu \nonumber\\ &\times&\left\{
\begin{tabular}{ll}
$f_{M}(m_{k} - \mu;M_{0},\sigma_{M})$ & (ITF)
\\ $1$ & (DTF)
\end{tabular}
 \right.
 \end{eqnarray}
The very interesting feature, which is shown by Eq.\,(\ref{BasicDist1}), is
that the distance modulus
estimate given by the DTF model, see Eq.\,(\ref{RoughDistanceEstimate}), does
not depend on the
luminosity distribution function of sources. Secondly, (in all cases) the
distance modulus estimate
given in Eq.\,(\ref{AHDistanceEstimate}) must be corrected for a bias. In order
to calculate the
correction term, we have to specify the luminosity distribution function
$f_{M}(M)$ in the case of
the ITF model, and the function $\kappa(\mu)$ in both cases. So let us assume
that the sources are
uniformly distributed in space ($h_{2}$), and eventually that they show a
Gaussian luminosity
distribution function ($h_{4}$). Hence, by using ({\it Def.}2.c,d), we obtain
straightforwardly
the following distance modulus estimate
 \begin{equation}\label{DistanceEstimate}
\mu_{0}^{(k)} = \left\{\begin{tabular}{ll}
$\frac{1}{1+\gamma^{2}}
\left(\left(\tilde{\mu}_{k}+\beta\sigma_{\zeta}^{2}\right)
+ \gamma^{2}(m_{k} - M_{0}) \right)$ & (ITF) \\
 $\tilde{\mu}_{k}+\beta\sigma_{\zeta}^{2}$ & (DTF)
\end{tabular}
 \right.,
 \end{equation}
where $\gamma=\gamma^{\rm ITF}$, see Eq.\,(\ref{SigmaZetaInequal}), with a
(accuracy) standard
deviation $\sigma_{\mu}^{(k)} = \sigma_{\mu}$,
 \begin{equation}\label{DistanceEstimateAccuracy}
\sigma_{\mu} = \left\{\begin{tabular}{ll}
$\sigma_{\zeta}/\sqrt{1+\gamma^{2}}$ & (ITF) \\
$\sigma_{\zeta}$ & (DTF)
\end{tabular}
 \right.
\end{equation}
According to Eq.\,(\ref{SigmaZetaInequal},\ref{zetaITF},\ref{SigmaZetaComp}),
it turns out that
these distance modulus estimates have similar accuracy, while information is
used for calculating the
ITF\,distance modulus estimate. By using
Eq.\,(\ref{aComp}-\ref{SigmaZetaComp}), we easily calculate the
difference $\Delta\mu = \mu^{\rm ITF}-\mu^{\rm DTF}$ between the distance
modulus estimates. We obtain
 \begin{equation}\label{DistanceDiff}
\Delta\mu = \frac{\gamma^{2}}{1+\gamma^{2}}\left(a^{\rm ITF}\langle p
\rangle_{1} + b^{\rm ITF} + \beta(\Sigma_{1}(M))^2 - M_{0}\right).
\end{equation}
Let us emphasize that $a^{\rm ITF}\langle p \rangle_{1} + b^{\rm ITF} +
\beta(\Sigma_{1}(M))^2$ is
an unbiased statistics which gives $M_{0}$ within $\sigma_{M}/\sqrt{N_{1}}$.
Therefore,
Eq.\,(\ref{DistanceDiff}) shows that the difference turns out to be a tiny
quantity. Namely,
$\Delta\mu$ has a vanishing expected value, with a standard deviation given by
 \begin{equation}\label{StDevDistanceDiff}
\sigma_{\Delta\mu}= \frac{\sigma_{\zeta}^{\rm
ITF}}{\sqrt{N_{1}}}\frac{\gamma}{\sqrt{1+\gamma^2}},
 \end{equation}
where $\gamma=\gamma^{\rm ITF}$, see Eq.\,(\ref{SigmaZetaInequal}). Therefore,
this shows that we
obtain the same distance modulus estimate by using different models.

Finally, we come to the conclusion that the choice of the model should be based
on the
reliability of hypotheses used about the selection effects, and it is
interesting to note
that the DTF approach is more robust than the ITF one. The correction terms in
Eq.\,(\ref{DistanceEstimate}) are not related to biases of Malmquist type but
identify to
{\it volume corrections}, herein calculated for {\it homogeneous} spatial
distributions or of
power law type (i.e., $\beta\neq\frac{3\ln10}{5}$), the {\it inhomogeneous}
case is
straightforward.

Obvious calculations show that the effects of calibration errors on distance
modulus estimates
make them less accurate by introducing a white noise of a $p$-dependent
standard deviation given by
\begin{equation}\label{DistanceStDCE}
\sigma^{\rm cal}_{\mu}=\sqrt{\sigma^{2}_{\delta_{a}}p^{2}-2{\rm
Cov}(\delta_{a},\delta_{b})p+\sigma^{2}_{\delta_{b}}},
\end{equation}
where $\delta_{a}$ and $\delta_{b}$ denote the calibration errors. The
simulations show that
\begin{eqnarray}\label{DistanceStDCEp}
\sigma_{\delta_{a}}= \left\{\begin{tabular}{ll} $0.21$ & (ITF) \\ $0.22$ &
(DTF)
\end{tabular}
 \right. \\
 \sigma_{\delta_{b}}= \left\{\begin{tabular}{ll} $0.53$ & (ITF) \\ $0.56$ &
(DTF)
\end{tabular}
 \right. \\
 {\rm Cov}(\delta_{a},\delta_{b})= \left\{\begin{tabular}{ll} $-0.11$ & (ITF)
\\ $-0.12$ & (DTF)
\end{tabular}
 \right.
\end{eqnarray}

Now it is natural to make the link between the distance modulus and the
$H_{0}$\,estimates. A
simple formal comparison between
Eq.\,(\ref{H0Inverse},\ref{H0Direct},\ref{DistanceEstimate}),
provides us with
\begin{equation}\label{H0Distance}
\frac{1}{N_{2}}\sum_{k=1}^{N_{2}}(\eta_{k}-\mu_{0}^{(k)}) =
\left\{\begin{tabular}{ll} ${\cal H}^{\rm ITF}$ & (ITF) \\
${\cal H}^{\rm DTF}$ & (DTF) \end{tabular}
 \right.
\end{equation}
It is obvious that such an equality is valid within the common set of
hypotheses, which confines
to those specified for
Eq.\,(\ref{H0Inverse},\ref{H0Direct},\ref{DistanceEstimate}). Now, we can
understand that the $H_{0}$\,statistics given in Eq.\,(\ref{H0InverseRem}) has
its foundation in a
context of distance estimates.

\section{Conclusion}\label{Conclusion}
We present a general framework to estimate the Hubble constant, as well as the
distances
of galaxies, when their peculiar velocities are neglected, by using distance
estimators
given by the Tully-Fisher, or the Faber-Jackson relations. Such relations can
be regarded
as a single law describing the observed linear correlation between the absolute
magnitude
$M$ of galaxies and their line width distance indicator $p$. This well known
problem has
been enlightened by taking into account a random variable $\zeta$ of zero mean
which
accounts for an intrinsic scatter of the TF\,relation ($M=a.p+b-\zeta$). The
method
consists of two steps~: the a priori choice of a statistical model, which is
defined
essentially on working hypotheses about the data distributions; and the
derivation of
parameters statistics by means of the maximum likelihood technique. This method
has the
advantage of providing unbiased estimates of model parameters, as long as the
selection
effects are taken into account by the statistical model. As standard, we assume
a
magnitude limited (complete) sample of uniformly distributed sources in space
which shows
a gaussian luminosity distribution function, although this method can easily be
extended
to more realistic situations. It turns clear that the presence of $p$-selection
effects
(which is not investigated here) makes this problem much more difficult,
although some
results require even weaker hypotheses.

We show that the ``Direct Tully-Fischer" and the ``Inverse Tully-Fischer"
methods identify as
maximum likelihood statistics related to particular models (herein, denoted ITF
and DTF), whose
difference limits on describing the TF~diagram in a different way. At first
glance, one might
wonder whether such an a priori choice is justified since these models replace
the one which should
be prescribed by the physics of galaxies (responsible for the $M$--$p$
correlation), and which is
not yet known. Fortunately, it is reassuring to point out that the estimates of
galaxies distances
and $H_{0}$ are not model dependent, contrarily to calibration parameters $a$
and $b$. Actually,
these models belong to a wide class of models, and both of them can be
interpreted as a choice of a
particular ``orientation'' for fitting the TF relation (according to usual
definitions). However,
the advantage of using models instead of fitting approaches is that one avoids
subjective
interpretations, for having clear-cut and unambiguous results. For example, we
easily understand
that, in order to obtain meaningful estimates, the calibration of the
TF~relation and the
determination of $H_{0}$, or the distances of galaxies, has to be performed
within the same model,
regardless of selection effects. Moreover, it turns out that the
$H_{0}$~statistics are still valid
when additional selection effects (or sampling rules) are present, which
informs us on the
robustness of these statistics. For example, in the case of the ITF~model,
selection effects with
respect to distance modulus, or redshift, and/or $M$, do not perturb the
estimate. On the other
hand, in the case of the DTF~model only additional selection effects with
respect to the redshift
are allowed.

The main result which ends the well known debate is that the ITF and DTF
estimates show identical
expectancies. Namely, the difference of estimating $H_0$, resp. a distance
modulus $\mu$, by mean
of ITF or DTF statistics (considered as a random variables) have vanishing mean
values and a
standard deviation of order of $\sigma_{\zeta}\gamma\sqrt{1/N_{\rm cal}+1/N}$,
resp.
$\sigma_{\zeta}\gamma\sqrt{1/N_{\rm cal}}$, where $N_{\rm cal}$ is the size of
the calibration
sample, $N$ is the size of the sample used to determine $H_{0}$, and where the
ratio
$\gamma=\sigma_{\zeta}/\sigma_{M}$ informs us on the gain of accuracy when
using the TF~diagram. In
practice, they are different only because of statistical fluctuations. It is
interesting to point out
that these approaches provide with us the same $H_{0}$~estimates when the
calibration sample and the
sample used to estimate $H_{0}$ show identical $p$-averages (herein called
``${\cal C}_{p}$-{\it
criteria}'').

Therefore, the choice between the ITF and the DTF approaches should be
motivated by
arguments about selection effects, accuracy and robustness of estimates.
Actually, in the
case of $H_{0}$~estimates, only the first criterion intervenes since the ITF
and the DTF
approaches show identical accuracy and robustness. With this in mind, we
introduce a
newly defined $H_{0}$~statistics, whose related model (herein denoted by
ITF$^\star$)
includes the ITF~model, where no hypothesis is required on the luminosity
distribution
function of sources, on their spatial distribution, and it is still valid when
the sample
is not complete. While it is a little less accurate (by a factor of
$\sqrt{1+\gamma^{2}}$), its advantage is to be much more robust than the ones
related to
the ITF and the DTF models. Finally, simulations show that $H_{0}$ can be
estimated with
an accuracy of the order of 5\% ($1\sigma$), which takes into account
calibration and
measurement errors (actually the first ones prevail on the other ones). In the
case of
distances, it turns out that the DTF estimate is more robust than the ITF
estimate,
because it does not depend on the luminosity distribution of sources. Both
estimates show
a correction for a bias, inadequately believed to be of Malmquist type.

\appendix
\section{Notations and useful formulas}\label{Notations}
The mathematical formalism is similar to the one used in Bigot \& Triay
(\cite{BigTri90a}). The
following features are addressed throughout the text by using the symbol ``{\it
Def.}''.
\begin{enumerate}
\item[{\it Def.}1] \label{pdfDef} The {\it probability density} ({\it pd}) of a
random
variable $x$ reads $dP(x)=f(x)dx$, where $f(x)$ represents the {\it pd
function} ({\it pdf}), we have $\int dP(x) = 1$. Sometimes, it is useful to
exhibit the {\it model parameters} involved in the statistical model, as the
mean $x_0$ and the standard deviation $\sigma$, by writing $f(x;x_0,\sigma)$.
\begin{enumerate}
\item \label{pdfDefGaussian} $g_{\rm G}(x;x_0,\sigma) =
(\sigma\sqrt{2\pi})^{-1}\exp{-\left((x-x_0)^2/(2\sigma^2)\right)}$ is a
Gaussian {\it
pdf}. \item \label{pdfDefNormal} A normal {\it pdf} can be written $g_{\rm
N}(x)=g_{\rm G}(x;0,1)$.
\item \label{pdfDefCumulNormal} The cumulative Normal {\it pdf} reads
${\cal N}(x) = \int_{-\infty}^{x}g_{\rm N}(t)dt$.
\end{enumerate}
\item[{\it Def.}2]\label{UsefullFormulas} Let $f$ be a {\it pdf}, and $\lambda$
be a
scalar value, in most of calculations, we use the following properties~:
 \begin{enumerate}
 \item \label{UFtranslation} $f(x+\lambda;x_0,\sigma)=f(x;x_0-\lambda,\sigma);$
 \item \label{UFDilation}
$f(\lambda
x;x_0,\sigma)=\lambda^{-1}f(x;\frac{x_0}{\lambda},\frac{\sigma}{\lambda})$;
 \item \label{UFGaussian1} $\exp(\lambda x)g_{\rm G}(x;x_0,\sigma)=
\exp \left(\lambda (x_0 + \lambda\frac{\sigma^2}{2})\right) g_{\rm G}(x;x_0 +
\lambda\sigma^2,\sigma)$.
 \item \label{UFGaussian2}$g_{\rm G}(x;x_1,\sigma_1) g_{\rm G}(x;x_2,\sigma_2)
=
 g_{\rm G}(x;x_0,\sigma_0) g_{\rm G}(x_1;x_2,\acute{\sigma})$,
 where $\acute{\sigma}=\sqrt{\sigma_1^2+\sigma_2^2}$, $x_0$ and $\sigma_0$ are
 defined as follows $\sigma_0^{-2} = \sigma_1^{-2}+\sigma_2^{-2}$ and
$x_0\sigma_0^{-2} =
 x_1\sigma_1^{-2} + x_2\sigma_2^{-2}$.
 \end{enumerate}
\item[{\it Def.}3]\label{Expectation} $P(h)=\int h(x)dP(x)$ denotes the
expected value
of the function $h(x)$.
\item[{\it Def.}4]\label{LikelihoodFunctionDef} The {\it pd} of a sample data
$\left\{{\cal G}_{k}\right\}_{k=1,N}$, which consists of $N$ independently
selected objects ${\cal G}_{k}$, is given by $\prod_{k=1}^{N}dP({\cal
G}_{k})$.
 \begin{enumerate}
\item \label{LikelihoodFunction} Its {\it pdf}, written in terms of {\it
 observables} (the measurable
random variables), but regarded as a function of model parameters, provides us
 with the {\it
likelihood function}.
\item \label{ML} ({\it The {\sc ml} method}.) The model parameters
statistics are obtained by maximizing the likelihood function, or
(equivalently) the natural logarithm of the {\it efficient part} of it, in
which the terms which do not contribute to the determination of parameters
are removed, herein briefly denoted by {\it lf}.
\end{enumerate}
\item[{\it Def.}5]\label{Statistics} We use the following usual definitions~:
 \begin{enumerate}
\item \label{Average}$\langle x \rangle=\sum_{k=1}^N x_{k}/N$ is the average,
\item \label{Covariance}${\rm Cov}(x,y) = \sum_{k=1}^N (x_{k}- \langle x
\rangle)(y_{k}- \langle y \rangle)/(N-1)$ is the covariance,
\item \label{StandardDeviation}$\Sigma(x) = \sqrt{{\rm Cov}(x,x)}$ is the
 standard deviation,
\item \label{CorrelationCoefficient}$\rho(x,y) = {\rm
Cov}(x,y)/(\Sigma(x)\Sigma(y))$ is the
correlation coefficient.
 \end{enumerate}
\item[{\it Def.}6]\label{Bias} The problem of biases in Statistics Theory is
well established~: an
{\it estimator} (or {\it statistic}) is biased when its expected value does not
correspond to model
parameter for which it has been made up. In practice, a bias is expected when
the normalization
factor depends on the model parameter, see Bigot \& Triay
(\cite{BigTri90a},\cite{BigTri90b}). For
instance, the average of absolute magnitudes, as provided by a sample of
objects brighter than a
given limiting apparent magnitude, is a biased estimator of the mean intrinsic
magnitude (that
characterizes the population of sources). Herein, such a bias is designated as
{\it bias of
Malmquist type}, a definition which can be extended to any bias due to
selection effects.
\item[{\it Def.}7]\label{Accuracy} The accuracy of an estimator is formally
defined as the reciprocal of its variance (The smaller the dispersion, the
greater the precision.).
\end{enumerate}

\section{Calculations involved in the ITF model}\label{CalculationsITF}
According to Eq.\,(\ref{BasicTh},\ref{BasicObs},\ref{InverseTFDensity}), the
normalization factor
reads
 \begin{equation}\label{ITFSNormalFactor}
	P_{\rm th}(\phi_{m}) = \int \phi_{m}(M + \mu)~f_{M}(M;M_{0},\sigma_{M})dM
 ~\kappa(\mu)d\mu,
 \end{equation}
which shows that it does not depend on parameters $a$, $b$ and $H_{0}$. Let us
note that if we
specify the functions $\phi_{m}$, $\kappa$ and $f_{M}$, according to hypotheses
($h_{1}$,$h_{2}$,$h_{3}$,$h_{4}$), see
Eq.\,(\ref{MagnitudeLimited},\ref{uniform},\ref{TFClass},\ref{MGaussian}), then
the normalization
factor is given by
 \begin{equation}\label{ITFNormalFactor}
	P_{\rm th}(\phi_{m}) \propto \exp{\beta\left(m_{\rm
lim}-M_{0}+\frac{\beta}{2}\sigma_{M}^2\right)}.
 \end{equation}

 \subsection{Calibration statistics}\label{CalibrationITF}
According to
Eq.\,(\ref{ApparentMagnitude},\ref{TFrelation},\ref{Rough},
\ref{DataCalibration},\ref{InverseTFDensity}),
the {\it pd} given in Eq.\,(\ref{BasicObs}) reads in terms of observables as
 \begin{eqnarray}\label{BasicCal}
	dP_{\rm obs} &= &\frac{\phi_{\rm m}(M + \mu)}{P_{\rm th}(\phi_{m})}
f_{M}(M;M_{0},\sigma_{M})dM ~\kappa(\mu)d\mu
 \nonumber \\ && \times g_{\rm G}(a~p+b;M,\sigma_{\zeta}) a\,dp,
 \end{eqnarray}
where the first right hand term is independent of $a$ and $b$, see
Eq.\,(\ref{ITFSNormalFactor}). Therefore, the {\it lf} ${\cal L}^{\rm ITF}_{\rm
cal}(a,b,\sigma_{\zeta})$ can be written as follows
 \begin{equation}\label{LikelyFunctionITF}
	{\cal L}^{\rm ITF}_{\rm cal} = \ln a - \ln \sigma_{\zeta}
-\frac{1}{N_{1}}\sum_{k=1}^{N_{1}}\frac{(a.p_{k}+b-M_{k})^2}
{2\sigma_{\zeta}^2}.
 \end{equation}
Hence, the {\sc ml} equations (obtained
by equating the partial derivatives of ${\cal L}^{\rm ITF}_{\rm cal}$ with
respect to $a$, $b$, and $\sigma_{\zeta}^2$ to zero) reads
 \begin{eqnarray}
 \label{LEquaITFa} a \langle p(ap+b-M) \rangle_{1} &=& \sigma_{\zeta}^2,\\
 \label{LEquaITFb} a \langle p \rangle_{1} + b &=& \langle M \rangle_{1},\\
 \label{LEquaITFc} \langle (ap+b-M)^2 \rangle_{1} &=& \sigma_{\zeta}^2.
 \end{eqnarray}
By expanding Eq.\,(\ref{LEquaITFc}) as $a\langle p(ap+b-M)\rangle +
b\langle(ap+b-M)\rangle-\langle M(ap+b-M) \rangle = \sigma_{\zeta}^2$. Hence,
it
follows that~:
\begin{itemize}
\item According to Eq.\,(\ref{LEquaITFa}), the first left hand term is equal to
 $\sigma_{\zeta}^2$.
\item According to Eq.\,(\ref{LEquaITFb}), the second left hand term is zero.
\item Thus we have $a\langle pM \rangle + b\langle M \rangle = \langle M^2
\rangle$. \item Hence, by subtracting $\langle M \rangle \times
(\ref{LEquaITFb})$, one
 obtains Eq.(\ref{aITF}).
\end{itemize}
Equation (\ref{bITF}) follows immediately from Eq.\,(\ref{LEquaITFb}). By
subtracting $a\langle p \rangle \times (\ref{LEquaITFb})$ to
Eq.\,(\ref{LEquaITFa}), we obtain $a^2(\langle p^2 \rangle-\langle p
\rangle^2) - a(\langle pM \rangle-\langle p \rangle\langle M
\rangle)=\sigma_{\zeta}^2$, which gives Eq.\,(\ref{zetaITF}).

\subsection{Determination of $H_{0}$}\label{DeterminationITF}
According to
Eq.\,(\ref{ApparentMagnitude},\ref{TFrelation},\ref{Rough},
\ref{DataCalibration},\ref{InverseTFDensity}),
the {\it pd} given in Eq.\,(\ref{BasicObs}) reads in terms of observables
$x$, $y$ and $\eta$, see Eq.\,(\ref{eta},\ref{variableX},\ref{variableY}),
as follows
 \begin{eqnarray}
 \label{BasicDet}
	dP_{\rm obs} &=& \frac{\phi_{m}(x+\eta)}{P_{\rm th}(\phi_{m})} f_{M}(x +
{\cal
H};M_{0},\sigma_{M}) \nonumber \\ && \kappa(\eta - {\cal H}) dx d\eta \times
g_{\rm G}(y;{\cal
H},\sigma_{\zeta})dy.
 \end{eqnarray}
It is important to note that this {\it pd} reads as a product of two
independent {\it pd}s, and
thus that
\begin{quote}
the distribution of the random variable $y$ does not depend on $x$ and $\eta$,
whatever the form of functions $f_{M}$, $\kappa$ and $\phi_{m}$.
 \end{quote}
The integration over $x$ and $\eta$ yields
\begin{equation}\label{H0densityITFS}
dP_{y}^{\rm ITF^\star} = g_{\rm G}(y;{\cal H},\sigma_{\zeta})dy,
\end{equation}
which shows that the expected value of $y$ provides us with $P_{\rm obs}(y)
= {\cal H}$. Therefore, the Hubble constant can be estimated by means
of the statistic given in Eq.\,(\ref{H0InverseS}). Moreover
Eq.\,(\ref{H0densityITFS}) shows that the standard deviation of the
$y$-distribution is equal to $\sigma_{\zeta}$. Thus, the standard deviation of
the
statistic providing $H_0$ is given by Eq.\,(\ref{ITFSAccuracy}).

If we specify the functions $\kappa$ and $f_{M}$, then we can perform the {\sc
ml} technique. By
assuming hypotheses ($h_{2}$,$h_{3}$,$h_{4}$), see
Eq.\,(\ref{uniform},\ref{TFClass},\ref{MGaussian}), the {\it lf} is given by
 \begin{eqnarray}\label{LikelyFunctionITFH0}
	{\cal L}^{\rm ITF}_{\rm det}(H_{0}) &=&
 -\frac{1}{N_{2}}\sum_{k=1}^{N_{2}}\frac{(y_{k}-{\cal H})^2}{2\sigma_{\zeta}^2}
- \beta {\cal H}
\nonumber \\ &-& \frac{1}{N_{2}}\sum_{k=1}^{N_{2}}\frac{(x_{k}+{\cal
H}-M_{0})^2}{2\sigma_{M}^2},
 \end{eqnarray}
since the normalization factor does not depends on $H_{0}$. Hence, the
likelihood equation ($d{\cal
L}^{\rm ITF}_{\rm det}/dH_0 = 0$) provides us with the statistic given in
Eq.\,(\ref{H0Inverse}).

In order to estimate the accuracy of the statistic (\ref{H0Inverse}), we
have to calculate the {\it pdf} of the following random variable
 \begin{equation}\label{H0InverseAux}
		z = \frac{ \sigma_{M}^2 y + \sigma_{\zeta}^2 \left( \left( M_{0} -
 \beta\sigma_{M}^2 \right) - x
\right) } { \sigma_{M}^2 + \sigma_{\zeta}^2 }.
 \end{equation}
Obvious calculations\footnote{- Eq.\,(\ref{BasicDet}) is written in terms of
variables $x$, $\eta$ and $z$, accordingly to
Eq.\,(\ref{MagnitudeLimited},\ref{uniform},\ref{MGaussian}), - one integrates
over $\eta$, and hence over $x$, we use
({\it Def.}2.c,d).} give
 \begin{equation}\label{H0densityITF}
dP_{z}^{\rm ITF} = g_{\rm G}(z;{\cal
H},\frac{\sigma_{\zeta}}{\sqrt{1+\gamma^2}})dz,
\end{equation}
see Eq.\,(\ref{SigmaZetaInequal}). According to
Eq.\,(\ref{zetaITF},\ref{ITFSAccuracy},\ref{H0densityITF}), and since
$\sigma_{M}\approx\Sigma_{1}(M)$, we obtain the accuracy given in
Eq.\,(\ref{ITFAccuracy}).

\section{Calculations involved in the DTF model}\label{CalculationsDTF}
Now, according to
Eq.\,(\ref{BasicTh},\ref{ApparentMagnitude},\ref{BasicObs},
\ref{Rough},\ref{DataCalibration},\ref{DirectTFDensity}),
it turns out that the random variables $M$, $p$ and $\mu$ are correlated
together\footnote{- $M$ and $p$, because of the TF diagram, - $M$ and
$\mu$, because the selection function $\phi_{m}(M + \mu)$ does not split into a
product
of two functions, - $\mu$ and $p$, as a consequence of above correlations.}.
Hence, the normalization
factor $P_{\rm th}(\phi_{m})$ becomes dependent on model parameters $a$ and
$b$. Thus, for proceeding
with the {\sc ml} technique, we have to calculate explicitly $P_{\rm
th}(\phi_{m})$ and its
derivatives with respect to model parameters, which forces us to presume a
priori the form of
functions $\phi_{m}(m)$, $f_{p}(p;p_{0},\sigma_{p})$ and $\kappa(\mu)$. We
assume
($h_{1}$,$h_{2}$,$h_{3}$,$h_{4}^\prime$), see
Eq.\,(\ref{MagnitudeLimited},\ref{uniform},\ref{pdistribution}). Hence, after
little
algebra\footnote{The calculation is straightforward by means of by part
integrations,
successively over $\mu$, $M$, and finally $p$, where ({\it Def.}2.c) is used
twice.},
it turns out that the normalization is still given by
Eq.\,(\ref{ITFNormalFactor}), with
\begin{eqnarray}
\label{DTFNormalFactorM0} M_{0} &=& ap_{0}+b, \\
\label{DTFNormalFactorSigmaM} \sigma_{M} &=&
\sqrt{a^2\sigma_{p}^2+\sigma_{\zeta}^2}.
\end{eqnarray}

\subsection{Calibration statistics}\label{CalibrationDTF}
According to
Eq.\,(\ref{ITFNormalFactor},\ref{DTFNormalFactorM0},
\ref{DTFNormalFactorSigmaM}),
which
shows that $P_{\rm th}(\phi_{m})$ depends indeed on parameters $a$, $b$,
$p_{0}$,
$\sigma_{p}$, and $\sigma_{\zeta}$, and to
Eq.\,(\ref{ApparentMagnitude},\ref{BasicObs},\ref{Rough},
\ref{DataCalibration},\ref{DirectTFDensity}),
the {\it lf} ${\cal L}^{\rm DTF}_{\rm
cal}(a,b,\sigma_{\zeta},p_{0},\sigma_{p})$ can be
written as follows
\begin{eqnarray}\label{LikelyFunctionDTF}
	{\cal L}^{\rm DTF}_{\rm cal} &=& -\ln
 P_{\rm th}(\phi_{m}) - \ln \sigma_{\zeta}
-\frac{1}{N_{1}}\sum_{k=1}^{N_{1}}\frac{(a.p_{k}+b-M_{k})^2}{2\sigma_{\zeta}^2}
\nonumber \\
 && - \ln \sigma_{p}
 -\frac{1}{N_{1}}\sum_{k=1}^{N_{1}}\frac{(p_{k}-p_{0})^2}{2\sigma_{p}^2}.
 \end{eqnarray}
According to Eq.\,(\ref{LikelyFunctionDTF}), the {\sc ml} equations
(obtained by equating the partial derivatives of ${\cal L}^{\rm DTF}_{\rm
cal}$ with respect to $a$, $b$, $p_{0}$, $\sigma_{p}^2$ and
$\sigma_{\zeta}^2$ to zero) can be written
\begin{eqnarray}
 \label{LEquaDTFa} \langle p(ap+b-M) \rangle_{1} &=& \beta\sigma_{\zeta}^2
\left(p_{0}-\beta\sigma_{p}^2a \right),\\
 \label{LEquaDTFb} a \langle p \rangle_{1} + b &=& \langle M \rangle_{1} +
 \beta\sigma_{\zeta}^2,\\
 \label{LEquaDTFp} p_{0} &=& \langle p \rangle_{1} + \beta\sigma_{p}^2 a,\\
 \label{LEquaDTFps} \langle (p-p_{0})^2 \rangle_{1} &=&
 \sigma_{p}^2\left(1+\beta^2\sigma_{p}^2 a^2\right),\\
 \label{LEquaDTFc} \langle (ap+b-M)^2 \rangle_{1} &=&
\sigma_{\zeta}^2\left(1+\beta^2\sigma_{\zeta}^2\right).
 \end{eqnarray}
Equations (\ref{LEquaDTFp},\ref{LEquaDTFps}) show that $\sigma_{p} =
\Sigma(p)$, while the Malmquist bias intervenes in the statistic Eq.
(\ref{LEquaDTFp}). According to Eq.(\ref{LEquaDTFp}),
Eq.\,(\ref{LEquaDTFa}) - $\langle p \rangle\times$Eq.\,(\ref{LEquaDTFb})
yields Eq.\,(\ref{aDTF}). By expanding $\langle (ap+b-M)^2 \rangle$,
according to Eq.(\ref{LEquaDTFc},\ref{aDTF}), we obtain
Eq.\,(\ref{zetaDTF}).

\subsection{Determination of $H_{0}$}\label{DeterminationDTF}
According to Eq.\,(\ref{HubbleL},\ref{ApparentMagnitude}),
the {\it pd} (\ref{BasicObs}) reads in terms of observables $x$, $y$ and
$\eta$, see Eq. (\ref{eta},\ref{variableX},\ref{variableY}), as follows
 \begin{eqnarray}
 \label{BasicDetH0Direct}
	dP_{\rm obs} &=& \frac{\phi_{m}(x+\eta)}{P_{\rm th}(\phi_{m})}
f_{p}(\frac{x+y-b}{a};p_{0},\sigma_{p}) \nonumber \\ && \kappa(\eta - {\cal H})
 \frac{1}{a} dx d\eta
\times g_{\rm G}(y;{\cal H},\sigma_{\zeta})dy
 \end{eqnarray}
Equations
(\ref{ITFNormalFactor},\ref{DTFNormalFactorM0},\ref{DTFNormalFactorSigmaM})
show that the
normalization factor $P_{\rm th}(\phi_{m})$ does not depend on $H_{0}$. Thus,
according to Eq.
(\ref{MagnitudeLimited},\ref{uniform},\ref{ZetaGaussian},\ref{pdistribution}),
 the {\it lf} reads
 \begin{equation}\label{LikelyFunctionDTFH0}
	{\cal L}^{\rm DTF}_{\rm det}(H_{0}) =
 -\frac{1}{N_{2}}\sum_{k=1}^{N_{2}}\frac{(y_{k}-{\cal H})^2}{2\sigma_{\zeta}^2}
- \beta {\cal H}
 \end{equation}
Hence, the likelihood equation ($d{\cal L}^{\rm DTF}_{\rm det}/dH_0 = 0$)
provides us with Eq.\,(\ref{H0Direct}). Obvious calculations\footnote{One
integrates the {\it pdf} given in Eq.\,(\ref{BasicDetH0Direct}) over $\eta$,
and
after over $x$, we use ({\it Def.}2.a,b,c), which gives
Eq.\,(\ref{ITFNormalFactor},\ref{DTFNormalFactorM0},
\ref{DTFNormalFactorSigmaM}).}
provide us with the {\it pdf} describing the distribution of the random
variable $y$,
 \begin{equation}\label{H0densityDTF}
dP_{y}^{\rm DTF} = g_{\rm G}(y;{\cal
H}+\beta\sigma_{\zeta}^2,\sigma_{\zeta})dy,
\end{equation}
which shows that the standard deviation is given by Eq.\,(\ref{DTFAccuracy}).

\section{Differences on the data description}\label{Difference}
In this section, we show that the {\rm DTF} model and the {\rm ITF} model
describe the data
distribution in a different way. This statement can easily be proved by
supposing the
antithesis, which is that the model parameters $a$ and $b$ are identically
defined in both
models (and thus also the random variable $\zeta$). Indeed, if $a$ and $b$ are
the same in
both models, the luminosity distribution function $f_{M}$ can be calculated
according to
Eq.\,(\ref{LuminosityFunction}) but within the {\rm DTF} model, as given by
Eq.\,(\ref{DirectTFDensity}). Now, by writing the {\it pdf} $f_{p}$ according
to
Eq.\,(\ref{pFunction}), thus within the {\rm ITF} model, see
Eq.\,(\ref{InverseTFDensity}),
we obtain two integrals that we transpose for obtaining the following
compatibility condition\footnote{The approximation is obtained by expanding the
right hand term, see also ({\it Def.}2.d).}
\begin{eqnarray}
f_{M}(M) &=& \int f_{M}(t) g_{\rm
G}(t;M,\sqrt{2}\sigma_{\zeta})dt, \\ &\approx& f_{M}(M) + \sigma_{\zeta}^2
\partial^2_Mf(M)
\nonumber
\end{eqnarray}
 which cannot be achieved, while such a disagreement is not so
drastic as that if the luminosity distribution function varies weakly within
ranges of the order
of $\sqrt{2}\sigma_{\zeta}$.

\section{Biases due to measurement errors}\label{ErrorsCorrection}
In order to calculate the magnitude of biases related to measurement errors, we
have to calculate the normalization factor $P_{\rm
th}\left(P_{\epsilon}^{(s)}\left(\hat{\phi}_{m} \right)\right))$, see
Eq.\,(\ref{BasicError1}). It turns out that one needs to specify the function
$\kappa(\mu)$,
and thus we assume a uniform spatial distribution of sources, i.e., ($h_{2}$),
see
Eq.\,(\ref{uniform}). Hence, it is clear that the integrations over the
$\epsilon_{\lambda}$
give unity, excepted for the one over $\epsilon_{m}$, because of selection
effects, see
Eq.\,(\ref{SelectEffectErr}). The calculation becomes evident if we use the
dummy
variable $\tilde{\mu}=\mu+\epsilon_{m}$, so we have $\kappa(\mu) =
\kappa(\tilde{\mu})
\exp{\left(-\beta \epsilon_{m}\right)}$, and then by using ({\it
Def.}2.b,c) we obtain Eq.\,(\ref{NormaliErrors}).
Hence, Eq.(\ref{BasicError1}) transforms as follows
\begin{equation}\label{BasicError2}
	d\hat{P}_{\rm obs}^{(s)} =
\hat{\phi}_{m} \frac{dP_{\rm th}}{P_{\rm th}\left(\phi_{m}\right)} \times
\exp{\left(-\frac{1}{2}\left(\beta \sigma_{\epsilon_{m}}^{(s)}
\right)^{2}\right)} dP_{\epsilon}^{(s)},
 \end{equation}
where it becomes clear that the selection function $\hat{\phi}_{m}$ plays the
role of a
correlation function between the variables $m$ and $\epsilon_{m}$. If the
measurement
errors were known then one could restore the values of observables from
Eq.\,(\ref{errorsm}-\ref{errorseta}), and then use the {\sc ml} technique for
obtaining
genuine statistics. In such a case, according to Eq.\,(\ref{BasicError2}), and
because
one has necessarily $\hat{\phi}_{m}(m_{k}) = 1$ for all individual datum
$(k=1,N)$, one
understands that one still obtains identical statistics to the ones given by
Eq.\,(\ref{H0InverseS},\ref{H0Inverse},\ref{H0Direct}), where the errors are
ignored.
However, since (in practice) the measurement errors are not known, the
$\epsilon_{\lambda}$-dependent parts of these statistics are substituted by
their expected
value according to the {\it p.d.} given in Eq.\,(\ref{BasicError2}). Let us
proceed with
preliminary calculations. It is easy to show that
\begin{equation}\label{ExpectErr}
\hat{P}_{\rm obs}^{(s)}\left(\epsilon_{\lambda}\right) = \left\{
\begin{tabular}{ll}
$-\beta \left(\sigma_{\epsilon_{m}}^{(s)}\right)^{2}$ & if
$\epsilon_{\lambda}=\epsilon_{m}$ \\ $0$ & otherwise
\end{tabular}
 \right.,
\end{equation}
and
\begin{equation}\label{ExpectSigErr}
\hat{P}_{\rm obs}^{(s)}\left(
\left(\epsilon_{\lambda} - \hat{P}_{\rm
obs}^{(s)}\left(\epsilon_{\lambda}\right)\right)^{2}
\right) = \left(\sigma_{\epsilon_{\lambda}}^{(s)}\right)^{2}.
\end{equation}
Accordingly to Eq.\,(\ref{ApparentMagnitude},\ref{errorsm},\ref{errorsmu}), for
{\it
Step}\,1, let us define the following variables
 \begin{eqnarray}\label{hatM}
 \hat{M} &=& \hat{m} - \hat{\mu} \\
 \epsilon_{M} &=& \epsilon_{m} - \epsilon_{\mu},
 \end{eqnarray}
so that the absolute magnitude reads
\begin{equation}\label{MagErrors}
 M = \hat{M} - \epsilon_{M}.
\end{equation}
Note that, because $\epsilon_{m}$ and $\epsilon_{\mu}$ are independent random
variables, we have
 \begin{equation}\label{MSigmaErrors}
 \left(\sigma_{\epsilon_{M}}^{(1)}\right)^{2} =
\left(\sigma_{\epsilon_{m}}^{(1)}\right)^{2} +
\left(\sigma_{\epsilon_{\mu}}^{(1)}\right)^{2}.
\end{equation}
Therefore, according to Eq.\,(\ref{ExpectErr},\ref{hatM}-\ref{MSigmaErrors}),
we
have
\begin{equation}\label{MErrors}
 \langle M \rangle_{1} = \langle \hat{M} \rangle_{1} +
\beta \left(\sigma_{\epsilon_{m}}^{(1)}\right)^{2},
\end{equation}
and since $M$ and $\epsilon_{M}$ are independent, it follows
\begin{equation}\label{SigmaMErrors}
 \Sigma_{1}^{2}(M) = \left(1-\delta_{M}\right)\Sigma_{1}^{2}(\hat{M}),
\end{equation}
see Eq.\,(\ref{DeltaErrorsM}). Similarly, it is evident to show that
 \begin{eqnarray}
\langle p \rangle_{1} &=& \langle \hat{p} \rangle_{1}, \label{pErrors} \\
(\Sigma_{1}(p))^{2}
&=& \left(1-\delta_{p}\right)(\Sigma_{1}(\hat{p}))^{2}, \label{SigmapErrors}
 \end{eqnarray}
see Eq.\,(\ref{DeltaErrorsp}), and thus that
\begin{equation}\label{RhoErrors}
 \rho^{2}_{1}(p,M) =
\frac{\rho^{2}_{1}(\hat{p},\hat{M})}{\left(1-\delta_{p}\right)
\left(1-\delta_{M}\right)}.
\end{equation}
For convenience in writing we use the following variable
 \begin{equation} \label{epsilonSigma}
\epsilon_{\sigma_{\zeta}}
 =\left(\frac{\delta_{p}}{1-\delta_{p}}+\delta_{M}\right),
\end{equation}
see Eq.\,(\ref{DeltaErrorsp},\ref{DeltaErrorsM}). In the case of the {\rm ITF}
model, according to
Eq.\,(\ref{aITF}-\ref{zetaITF},\ref{MErrors}-\ref{DeltaErrorsp},
\ref{epsilonSigma})
one has
 \begin{eqnarray}
 \label{aITFErrors}
a^{\rm ITF} &=& \hat{a}^{\rm ITF}\left(1-\delta_{M}\right), \\
 \label{bITFErrors}
b^{\rm ITF} &=& \hat{b}^{\rm ITF} - \delta_{M}\hat{a}^{\rm ITF}\langle \hat{p}
\rangle_{1}
+ \beta \left(\sigma_{\epsilon_{m}}^{(1)}\right)^{2} , \\
 \label{zetaITFErrors}
\left(\sigma_{\zeta}^{\rm ITF}\right)^{2} &=&
\left(1-\delta_{M}\right)\left(\hat{\sigma}_{\zeta}^{\rm ITF}\right)^{2}
 \nonumber\\&-&
\epsilon_{\sigma_{\zeta}}\left(1-\delta_{M}\right)\left(1-\delta_{p}
\right)\left
 (\hat{a}^{\rm ITF}\right)^{2}(\Sigma_{1}(\hat{p}))^2.
\end{eqnarray}
Hence, according to Eq.\,(\ref{variableY}), we easily obtain
Eq.\,(\ref{H0ITFSErrors}). Similarly, for the DTF model, we obtain
\begin{eqnarray}
 \label{aDTFErrors}
a^{\rm DTF} &=&
 \frac{\hat{a}^{\rm DTF}}{\left(1-\delta_{p}\right)}, \\
 \label{bDTFErrors}
b^{\rm DTF} &=&
 \hat{b}^{\rm DTF} + \delta_{M}\left(\hat{a}^{\rm DTF}\langle \hat{p}
\rangle_{1} -
\beta\left(\hat{\sigma}_{\zeta}^{\rm DTF}\right)^{2}\right) \nonumber \\
 &-& \epsilon_{\sigma_{\zeta}} \left(
\hat{a}^{\rm DTF}\langle \hat{p} \rangle_{1}
- \beta\left(\hat{\sigma}_{\zeta}^{\rm DTF}\right)^{2}
+ \beta(\Sigma_{1}(\hat{M}))^2\right) \nonumber \\ &+&
\beta\left(\sigma_{\epsilon_{m}}^{(1)}\right)^{2}\\
 \label{zetaDTFErrors}
\left(\sigma_{\zeta}^{\rm DTF}\right)^{2} &=&
 \frac{1}{1 - \delta_{M}}
\left(\hat{\sigma}_{\zeta}^{\rm DTF}\right)^{2} -
\epsilon_{\sigma_{\zeta}}(\Sigma_{1}(\hat{M}))^2.
\end{eqnarray}
Hence, according to Eq.(\ref{variableY}), we easily obtain
Eq.(\ref{H0DTFErrors}).

\acknowledgements
L. Gouguenheim is kindly thanked for enlightening comments, L. Bottinelli and
G. Paturel for
discussions. One of us (R.T) thanks E. Giraud and B. Tully for fruitful
discussions. Financial
support from {\it GdR Cosmologie et Grandes Structures } has been appreciated.

\end{document}